\documentclass[10pt,journal,letterpaper,compsoc]{IEEEtran}
\usepackage{url}
\usepackage{amsmath}
\usepackage{amssymb}
\usepackage{multirow}
\usepackage[usenames]{color}
\usepackage[pdftex]{graphicx}

\makeatletter
\def\url@leostyle{%
  \@ifundefined{selectfont}{\def\UrlFont{\sf}}{\def\UrlFont{\small\ttfamily}}}
\makeatother
\newcommand{\comment}[1]{}
\newcommand{\mybf}[1]{\textcolor{red}{#1}}
\newcommand{\myem}[1]{\textcolor{blue}{\bf #1}}

\begin{document}

\title{Securing Your Transactions: Detecting Anomalous Patterns In XML Documents}

\author{Eitan~Menahem,
        Alon~Schclar,
        Lior~Rokach,
				Yuval~Elovici% <-this % stops a space
\IEEEcompsocitemizethanks{
\IEEEcompsocthanksitem E. Menahem, L. Rokach and Y. Elovici are with the Department of Information Systems Engineering, Ben-Gurion University of the Negev, Be'er Sheva, 84105, Israel.\protect\\
E-mail: \{eitanme, liorrk, elovici\}@post.bgu.ac.il
\IEEEcompsocthanksitem A. Schclar is with the School of Computer Science, The Academic College of Tel-Aviv-Yafo, Tel Aviv, 61083, Israel.
E-Mail: alonschc@mta.ac.il}% <-this % stops a space
\thanks{}}

%\markboth{Transactions on Dependable and Secure Computing,~Vol.~X, No.~X, January~2013}%
%{Menahem \MakeLowercase{\textit{et al.}}: Securing Your Transactions: Detecting Anomalous Patterns In XML Documents}

%\IEEEpubid{\makebox[\columnwidth]{\hfill 0000--0000/00/\$00.00~\copyright~2007 IEEE}%

\IEEEcompsoctitleabstractindextext{%
\begin{abstract}
XML transactions are used in many information systems to store data and interact with other systems. Abnormal transactions, the result of either an on-going cyber attack or the actions of a benign user, can potentially harm the interacting systems and therefore they are regarded as a threat.  
In this paper we address the problem of anomaly detection and localization in XML transactions using machine learning techniques.  
We present a new XML anomaly detection framework, \emph{XML-AD}.   Within this framework, an automatic method for extracting features from XML transactions was developed as well as a practical method for transforming XML features into vectors of fixed dimensionality. With these two methods in place, the XML-AD framework makes it possible to utilize general learning algorithms for anomaly detection. Central to the functioning of the framework is a novel multi-univariate anomaly detection algorithm, ADIFA. The framework was evaluated on four XML transactions datasets, captured from real information systems, in which it achieved over 89\% true positive detection rate with less than a 0.2\% false positive rate.

%many types of anomaleis fall victim to one or more adversary attacks, including: dictionary and buffer overflow attacks, cross-site scripting, SQL injection, parameter tampering, information leakage and more. Many of these attack can be detected using anomaly detection techniques which assume that poisoned XMLs .
\end{abstract}

%\category{C.2.0}{General}{Security and protection}
%\category{K.4.4}{XML Transactions}{Security}
%\category{I.2.6}{Learning}{Concept and parameters Learning}

% Note that keywords are not normally used for peer review papers.
\begin{keywords}
XML Anomaly Detection, XML Security, Machine-Learning, Outliers Detection, Anomaly-Detection
\end{keywords}}

\maketitle

% is not selected <OR> if conference mode is selected - because compsoc
% conference papers position the abstract like regular (non-compsoc)
\IEEEdisplaynotcompsoctitleabstractindextext

% \fi
\IEEEpeerreviewmaketitle

\section{Introduction}
Today, many information systems communicate and interact through XML transactions. These transactions may fall victim to cyber-attacks or even benign mistakes, which can alter the structure and content of their interaction media, i.e., the XML documents.  Regardless of whether the origin of these alterations is malicious or benign, the altered XMLs, especially those that adhere to the XSD schema, can potentially exploit vulnerabilities of the interacting information systems. Since the altered XMLs can be render as anomalous with respect to the majority of the XML transactions in the same domain, detecting anomalous XMLs is an important means to increase the security of many information systems. Unfortunately, state-of-art, end-to-end security protocols for XML transactions, i.e., XML encryption \cite{xmlencryption02}, XML signature \cite{xmlsignature02}%, DamianiVS02}
, and XML-canonicalization \cite{xmlcanonicalization01} provide little protection against such a threat. This is mostly because the alteration actions, which deform the XMLs, take place before such protective measures are applied at the endpoint's systems.
It follows that in addition to the abovementioned security protocols, XML documents should be subjected to an anomaly detection prior to being processed by the endpoint information systems.

Extensible Markup Language \cite{dcd} is a framework that facilitates the definition of structured markup languages. Data in such languages is described by documents in which every datum is encapsulated by tags. Since XML is very flexible it is especially suitable for: (a) data storage in a structured format; (b) data transfer and sharing both in local organizational networks and over the Internet; and (c) data serialization. 
The XML files, bound to their definition in an XSD schema can vary considerably and two XMLs that adhere to the same XSD schema, can have very different attributes regarding both their content and structure.\\

\subsection{XML Anomalies}
Anomalies are data patterns, which are either very rare or novel. In the scope of this paper, the anomalous patterns are related to both the structure and content of an XML document. Such patterns can be generated by either actions of which intentions are malicious (i.e., cyber-attack), or benign (user mistake or a technical error). Next, we describe the two most prominent anomalous pattern generators.

\subsubsection*{XML attacks}
Applications that interact through XML messages, such as various Web-services, are essentially vulnerable to a wide range of malicious attacks. These attacks exploit various vulnerabilities in the XML processing mechanism, such as the vulnerability of XML parsers or the weak points of input verification in the target server application. %We refer to these types of attacks as text-based attacks. 
Among the prominent attacks of this type are input validation attacks \cite{Moradian06}; probing \cite{VorobievH06a}; malware infiltration; buffer overflow \cite{VorobievH06a, Moradian06}; XML parameter poisoning \cite{VorobievH06a,stamos2005attacking}; CDATA field attacks \cite{VorobievH06a,stamos2005attacking}; SQL injection \cite{VorobievH06a,stamos2005attacking,Moradian06}; cross-site scripting \cite{Moradian06}; schema poisoning \cite{lindstrom2004attacking}; denial of service (DoS); DDoS – XML bombardment; DOM parser DoS attacks; XML Bomb \cite{XMLBomb2009} and repetition attack. These XML attacks are expected to result in producing XML anomalies, since the attacks are expressed through string expressions (or by other data types) that, with respect to the \emph{normal} XML transaction collections, are inherently very unlikely to be obtained.

Another threat to modern information systems arises from data leakage. Among the causes of data leakage are Trojan attacks, SQL injection attacks, or simple human error. There are many ways in which outgoing XML transactions can lead to data leaks in the system.  The simplest way results from putting all the data, as it is, in one field that is not properly constrained by a regular expression. A simple variation of this scheme is dividing the data into several parts and embedding it into many different fields of the XML file. 

\subsubsection*{Benign Anomalies}
Not all XML anomalies are a product of a cyber-attack or a malicious action. There are many ways in which XML documents might become anomalous. User mistakes, application errors and communication errors are typical examples of how benign anomalies might be manifested in XMLs.

\subsection{Problem Statement and Applicability}
\label{problemStatement}
The present work focuses on the problem of detecting and localizing anomalies in readable XML documents at computer endpoints, but will not address XML parser attacks since those can be efficiently detected at the network level, by technologies such as \emph{Anagram} \cite{DBLP:conf/raid/WangPS06}.

The algorithms, which are presented in this paper aim to detect anomalies that stem from either malicious or benign actions. In our opinion, it is important to detect both types of anomalies. The rationale behind this approach is the assumption that XML anomalies, regardless of their nature, have the potential to result in unwanted effects in the information processing system. \\
We would like to stress that the present work does not try to infer the nature of the detected anomalies, since this require elaborate forensic work and an understanding of the data domain and semantics.%, which is beyond the scope of this work.
Consequently, we use \emph{XML-AD} only as an indicator for what could be a network or system attack, which is being delivered by XML documents. As such, \emph{XML-AD} can be applicable, for example, for endpoint anomaly-based XML-Firewalls.

\subsection{Anomaly Detection}
\label{sec:litratureSurvey}
Anomaly detection \cite{AgyemangBA06, Bakar06, Beckman1983, HodgeA04, MarkouS03a, PatchaP07} is a process aimed at discovering patterns in datasets that deviate from the behavior or the expect behavior of the majority of the data. Anomaly detection can be found in a broad spectrum of applications such as intrusion detection, cyber-security, fraud detection, financial systems, and military surveillance, to name a few. Anomaly detection methods employ a wide range of techniques that are based on statistics, classification, clustering, nearest neighbor search, information theory and spectral analysis. 

There might be an infinite number of anomalous patterns, some of which are very rare and hard to obtain. In such cases, the most conventional learning approach, i.e., supervised-learning, is impractical since training a supervised classifier demands at least a single example from each of the patterns that must be classified. Moreover, in many real-life domains, normal state examples are inherently easier to obtain than anomalous state examples. In such domains, researchers take the semi-supervised anomaly detection approach (also known as one-class learning), in which only the normal class is being taught but the algorithm learns to detect abnormal patterns \cite{HodgeA04}.\\
Semi-supervised anomaly detection is a suitable approach for training anomaly detection for XML transactions when one assumes that at the time of training, only normal XML examples exist.

\begin{figure*}[ht]
	\centering
		\includegraphics[scale=.6]{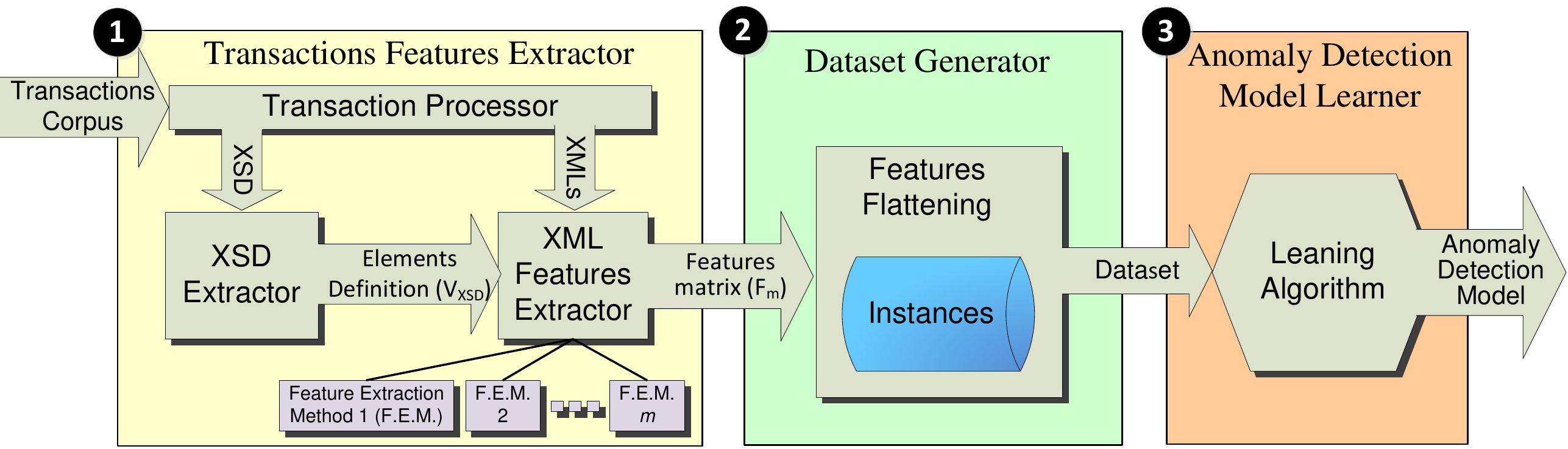}
	\caption{The conceptual architecture of XML-AD} 
	\label{fig:ADIFA_Process}
\end{figure*}

\subsection{Similarity-Based Anomaly Detection}
In order to classify a new XML example as \emph{normal} or as \emph{anomalous}, most anomaly detection algorithms consider the similarity of the new example to a group of XML documents, labeled \emph{normal}. The general idea is that if the new example is similar to the \emph{normal} instances, it should be labeled \emph{normal}; otherwise, it should be labeled \emph{anomalous}. Such a similarity function, denoted as $S(\cdot)$, takes two parameters: a new instance, $x_{new}$, and a group of XML documents, labeled as \emph{normal}, $X^{\{normal\}}$. One way to calculate such a similarity score is by computing the distance from $x_{new}$ to the closest \emph{normal} XML document, i.e., $S(x_{new},X)\equiv argmin_{x\in X^{\{normal\}}}d(x_{new},x)$, where  a distance function, ${d(\cdot)}$ %\mapsto\Re}$
, is a pairwise distance function. The distance function, $d(x_{new},x)$, computes a scalar that reflects the XMLs` mutual similarity in the feature space. Finding such a distance function, which is able to separate the \emph{normal} documents from the \emph{anomalous} documents in the feature space, is a fundamental challenge in XML anomaly detection \cite{LongSS05}.

Most similarity-based anomaly detection algorithms use a multivariate vector distance function, which is a plausible approach in many domains, especially those in which data is represented by low-dimensional vectors. However, multivariate distance functions are inherently susceptible to the "curse of dimensionality" \cite{bellman1957}. Consequently, the functions become much less accurate as the number of dimensions grows, since between any two points in the given dataset the distance converges, rendering the concept of distance meaningless. Another weak point one finds in similarity-based anomaly detection algorithms that use multivariate vector distance functions is that they are unable to indicate the specific dimension(s) in the new vector that incurs the anomalous pattern. As a result, the algorithms do not allow localization of the anomaly pattern source, i.e., detect the cause of the anomaly.

To avoid the weaknesses of the abovementioned similarity-based anomaly detection algorithms, the approach taken in this study is to use multiple, univariate distance functions. This approach was previously used in several domains, including intrusion detection \cite{%barbara2001detecting,
sebyala2002active,valdes2000adaptive,patcha2007overview,bronstein2001self}, however, it has not been applied to detect XML anomalies so far. We show that our approach results in a very accurate XML anomaly detection and makes it possible to localize the dimensions (i.e., the XML features), which are the source of the anomalous patterns.

\subsection*{Paper outline}
The rest of this paper is organized as follows. In Section \ref{RelatedWork} we present related work and discuss the need for a new anomaly detection framework for XML transactions. In Section \ref{XML-AD}  we present \emph{XML-AD}, the proposed anomaly detection framework. In Section \ref{methods} we discuss the methods, classifiers, datasets and performance metrics used for evaluating XML-AD. In Section \ref{evaluation}, we present several evaluation experiments and discuss their results. In Section \ref{conclusion}, we summarize the contributions of this paper and discuss future directions.

\section{Related Work}
\label{RelatedWork}
Despite the risk XML documents anomalies impose, very little relevant research has been done in this area. In the following paragraphs, we describe available methods for the anomaly detection of XML forms. All these methods take the semi-supervised anomaly detection approach.\\
Bruno et al.  propose a method for detecting frequently occurring relationships in datasets, which correspond to the normal behavior of the data \cite{BrunoGQR07a, BrunoGQR07b}. The detection method uses association rules and the relationships are represented as quasi-functional dependencies. Anomalies are discovered by querying either the original database or the previously mined association rules to indicate the presence of erroneous data or novel information that represents the outliers of frequent rules. The method is independent of the considered database and directly infers rules from the data. In \cite{BrunoGQR10}, an incremental approach is used to extend the method in \cite{BrunoGQR07a, BrunoGQR07b} to handle dynamic databases where the anomalies must be updated according to changes that the data undergoes.

Premalatha and Natarajan \cite{PremalathaN09} mine negative association rules \cite{WuZZ04}, which are used to describe those relationships between item sets that indicate the occurrence of some item sets by the absence of others. The chi-square test is used to identify independent attributes and the anomalies are identified as a negative association rule whose confidence value is greater than a minimum confidence threshold. Unfortunately, domain knowledge of the data sets is incorporated into filter rules, a step that does not contribute to the detection process.

H{\'e}v\'{\i}zi et al. \cite{HeviziML04} use probabilistic inference for classification and anomaly detection of structured documents which they test on XML documents. Specifically, they extract a feature vector from every XML document according to the number of attributes each tag can have. The features are learned and represented in a factorized form as a product of pairwise joint probability distribution functions according to a method introduced by Chow and Liu \cite{Chow68approximatingdiscrete}. Anomaly is detected by applying an acceptance threshold to the probability values. The authors indicate that this threshold should be trained and adapted for databases that are subject to frequent changes.

Raz et al. \cite{RazKS02} detect anomalies in dynamic data feeds. Specifically, invariants such as value interval and arithmetic expressions are extracted and used as proxies to detect anomalies. The detection method is demonstrated for semantic anomalies, i.e., values that are syntactically correct but have unreasonable values. Two types of invariants are extracted, namely, the mean statistics and invariants that are produced by an adjusted version of a software for the detection of invariants in computer programs. \\

In light of the above mentioned works, the original contribution of this work is three-fold: (1) it presents a general and automatic method for extracting structural and content features from XML transactions, (2) it provides a practical method for transforming XML features into vectors of fixed dimensionality and hence, enables the use of non-proprietary machine-learning algorithms for the XML anomaly detection task, (3) it presents a novel anomaly detection algorithm, ADIFA, which is based on the multi-univariate model approach.

\section{XML-AD: An Automated, Context-Learning of XML Documents}
\label{XML-AD}
We propose a new framework, \emph{XML-AD}, for training a classifier for detecting and localizing XML anomalies. The framework is composed of three stages: feature extraction, dataset generation and training of machine-learning model. The input for the training process is a corpus of XML transactions and a single XSD file, which defines the transactions at hand. \\
In the first stage, the raw transaction features are extracted. In order to do so, the XSD is parsed and the meta data it contains, i.e., XML elements definition and constraints, are extracted. Next, the entire training XML corpus is put through a feature extraction procedure. The meta-data extracted from the XSD is used to select the suitable feature extraction methods for each of the available XML elements. 
The second stage is dataset generation, in which the transaction features are aggregated and arranged in tuples, each containing multiple computed (aggregated), features of a single transaction. These tuples are then added to a dataset (the train-set). 
In the third and final stage, the anomaly detector is trained by applying the ADIFA algorithm (section \ref{sec:ADIFA}), on the train-set, which was generated in the previous stage. The above mentioned process is depicted in Figure \ref{fig:ADIFA_Process}.

\subsection{Transactions Feature Extraction}
\label{FeatureExtraction}
\begin{figure}[htb]
	\centering
		\resizebox{1\linewidth}{!} {
		\includegraphics{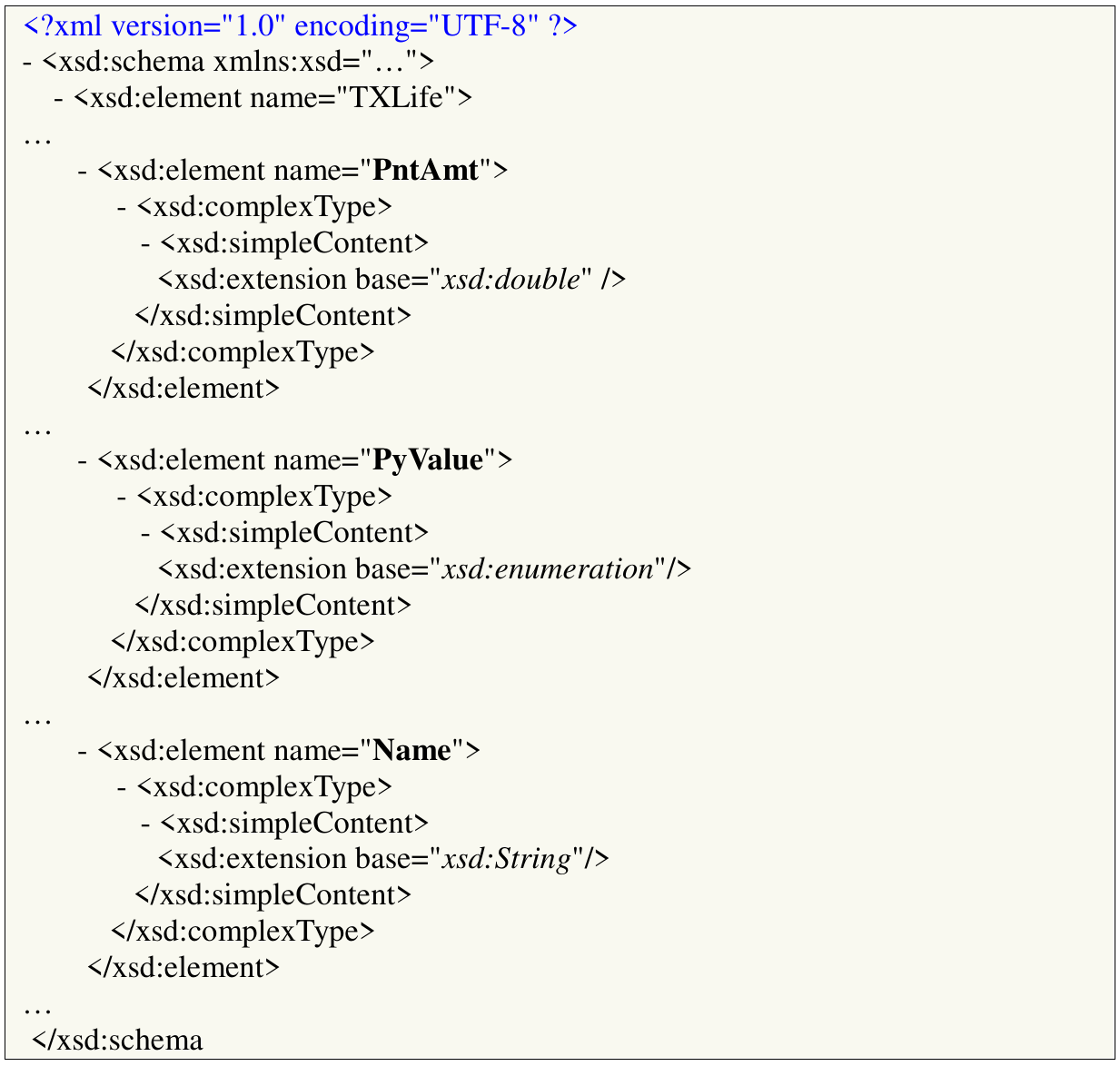}
		}
	\caption{An example of an XSD file}
	\label{fig:XSD}
\end{figure}

\begin{figure}[htb]
	\centering
		\resizebox{1\linewidth}{!} {
		\includegraphics{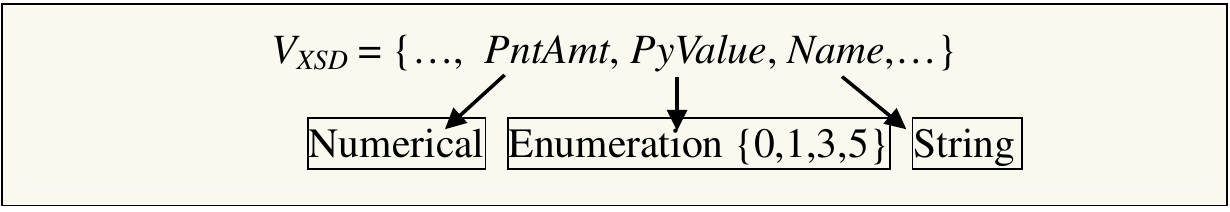}
		}
	\caption{A part of the XSD vector, $V_{XSD}$, produced by the XSD parser. Three objects are Visible: $PntAmt$, $PyValue$ and $Name$ of type Numerical, Enumeration and String, respectively.}
	\label{fig:XSD_vector}
\end{figure}

The feature extraction process starts with the acquisition of the definitions of the XML elements. This is achieved by parsing the XSD file. The XML elements' definitions are then stored in a data structure that we denote as $V_{XSD}$. 
In Figure \ref{fig:XSD} we show an XSD file that defines three variables: \emph{PaymentAmount}, \emph{PyValue} and \emph{Name}. The first variable is defined as \emph{XSD:double}, which means that a numerical value may be assigned to it. \emph{PyValue} was defined as \emph{enumeration}, so only a pre-defined (not visible in this example), integer may be assigned to it. Last, the \emph{Name} variable is defined as \emph{XSD:String} indicating that any textual symbols may be assigned to it. 
%XSD schema contains many built-in data types. The most common types are: \emph{xsd:string}, \emph{xsd:decimal}, \emph{xsd:integer}, \emph{xsd:boolean}, \emph{xsd:enumeration}, \emph{xsd:date} and \emph{xsd:time} encompassing strings, numeric, enumeration, time and encoded binary data types. 
In Figure \ref{fig:XSD_vector} we can see that $V_{XSD}$ contains one descriptive object for each defined element in the XSD file shown in Figure \ref{fig:XSD}. A descriptive object is a simple container that carries the type of the XML element (i.e. \emph{numeric}, \emph{date}, \emph{binary} etc.). In case of enumeration the descriptive object also contains value ranges (as with \emph{PyValue} above). To avoid the complexity of handling many XSD data-types, we found that it was sufficientto deal with only a few abstract data types: \emph{Numerical}, \emph{Enumeration}, \emph{String} and \emph{Date}.

Next, the XML transactions are processed and their features are extracted and stored in the \emph{features matrix} ($F_m$). The $F_m$ matrix will contain a single row for each XML transaction. Each row is comprised of multiple complex features, one for each element definition stored in $V_{XSD}$. Since the XML structure allows repetition of elements within the same document, a complex feature may contain multiple occurrences of the same element. Therefore, each complex feature contains a list, $\{mv_1, mv_2,\dots\}$, which we denote as 'measurement-vector' ($mv$), which stores information involving the occurrences of the related element. Each term in the 'measurement-vector' corresponds to a single element occurrence. Finally, each $mv$ contains $l$ scalar measurements, $\{m_1, m_2,\dots, m_l\}$ that corresponds to $l$ attributes of each XML element occurrence.

\begin{figure}[t]
	\centering
	\resizebox{1\linewidth}{!} {
		\includegraphics{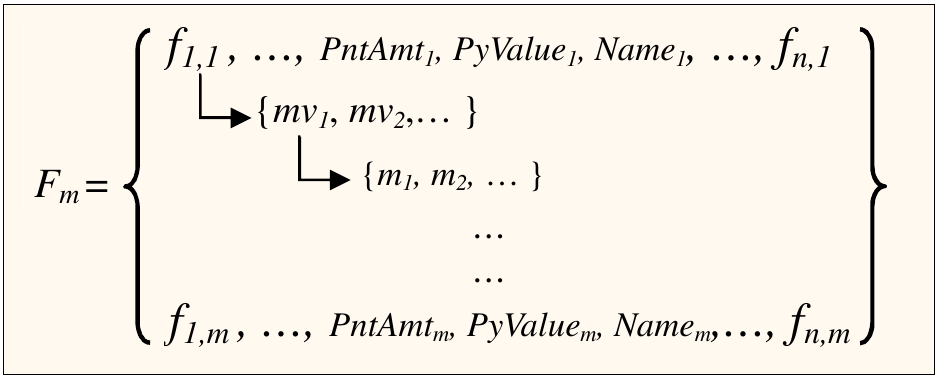}
		}
	\caption{The extracted features matrix, $F_m$. $m$ and $n$ are the number of transaction instances and the number of complex features, respectively.}
	\label{fig:FmMatrix}
\end{figure}

Figure \ref{fig:FmMatrix} exemplify the $F_m$ matrix related to the above XSD example.

\subsection{Dataset Compilation}
\label{datasetGeneration}
Traditionally, most machine-learning algorithms require that two conditions regarding their input datasets be met: (1) all instances must have the same number of features, and (2) all features must be scalars (as opposed to the complex features, which have an inner structure). While transaction instances in $F_m$ contain the same number of complex features, they may contain a different number of inner data items (e.g., measure-vectors). Thus, the second condition is not met.\\

To overcome this problem, $F_m$ should be flattened. The flattening process can be made in either a lossless or a lossy manner. 
In the lossless flattening process the instances information is preserved, however, the number of features per instance is not constant and consequently, the derived dataset will not be rectangular. Therefore, the first condition mentioned above is not met.
In the lossy flattening process, some information is sacrificed, while new properties can be gained, for example, determining the number of features per instance. In our case, the benefit of choosing the lossy flattening alternative is that many usable generic machine-learning algorithms can be used later to train anomaly detection models. 

\begin{figure}[b]

	\centering
	\resizebox{1\linewidth}{!} {
		\includegraphics{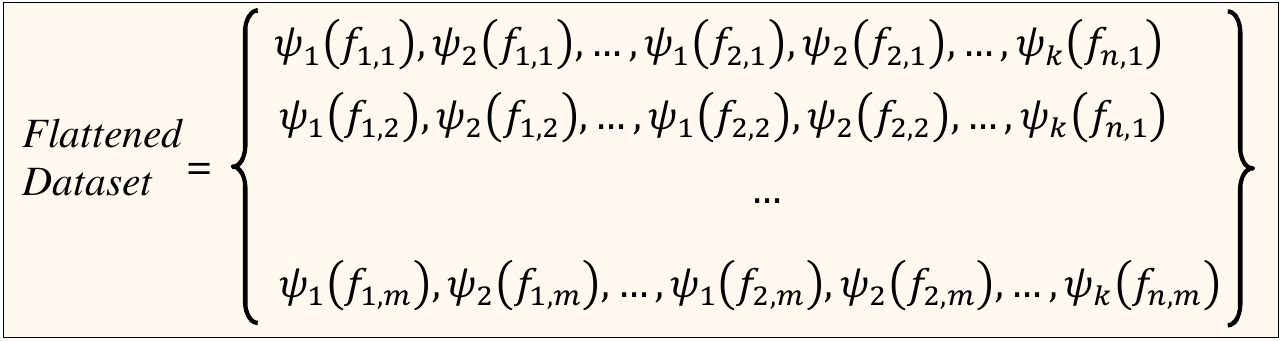}
		}
	\caption{The derived flattened dataset}
	\label{fig:flattened_dataset}
	
\end{figure}

We chose to flatten the $F_m$ matrix using a lossy process due to the above mentioned advantages. The flattening was achieved by applying a group of aggregate functions, $\Psi=\{\psi_1,\dots,\psi_k\}$, to each of $F_m$'s tuples separately, where $\psi_i: \{f_{1,\dots,n}\} \rightarrow \mathbb{R}$. Specifically, let $f_{i,j}$ be a complex feature where $i$ denotes the feature and $j$ denotes the tuple. A set of $k$ aggregated values $\psi_l (f_{i,j})$ is calculated where $l=1,\ldots,k$. Since the same aggregate functions are applied to all the tuples in $F_m$, the derived flattened dataset is rectangular, as Figure \ref{fig:flattened_dataset} shows. Next, we discuss the aggregate functions.

\subsection*{Aggregate Features}
Using aggregate functions to flatten the complex feature matrix, $F_m$, inevitably results in the loss of some information. However, to minimize the loss of information, the aggregate functions should only discard information that would have as little impact as possible on the ability to detect anomalies. 

Specifically, for \emph{Numerical} and \emph{Time} complex-features, we apply the \emph{Maximum} and \emph{Minimum} aggregation functions. Accordingly, only the maximal and minimal values of the complex feature’s data-times are used as simple (scalar), features.
For \emph{Enumeration} complex-features we use the \emph{Sum} aggregation function. By summing up every occurrence of each possible enumeration value, the number of generated simple-features becomes dependent on the number of elements in the enumeration. 
\emph{String} complex-features are treated slightly differently. First, we compute the number of words and the length of the string for each string occurrence in a String complex-feature. Then, we apply the \emph{Maximum} and \emph{Minimum} aggregate function over these values, to produce four simple features: the maximal and minimal number of words, and the maximal and minimal string length. Additionally, to learn the textual context of the XML transactions, a dictionary of words is produced from the training transaction, from which we then compute the TF-IDF (word term frequency inverse document frequency) value for every word in the dictionary. Lastly, according to their TF-IDF value, only the $k$ most prominent words are chosen for extracting simple features, i.e., TF-IDF values. This way, each transaction instance will contain $k$ simple text contextual features.\\

Finally, when the flattening process is over, the derived dataset meets both of the required conditions mentioned above: ALL instances have the same dimensionality and all the features are scalars. Evidently, as we illustrate in Section \ref{evaluation}, the proposed flattening approach was very effective for detecting XML anomalies.

\subsection{The Anomaly Detection Model}
\label{sec:ADIFA}
Due to the flattening process, a derived dataset has two essential properties that make it difficult to learn. First, its dimensionality is high, because the flattening process translates each XSD element into several simple features, and standard XSDs contain thousands of such elements. Moreover, since the learning task is \emph{anomaly detection},  its dimensionality cannot be reduced by feature selection because anomalies can appear in any of the dataset's features, thus, removing any feature can result in a missed anomaly detection.\\
In addition, since it is assumed that the dataset is composed of \emph{normal} instances only, the dataset represents a one-class classification problem, which is inherently a more difficult task compared to supervised or unsupervised anomaly detection. While there is wealth of anomaly detection algorithms to choose from, only very few are capable of a one-class classification, from which none, as far as we know, can cope with datasets whose dimensionality is very high.\\ To close this gap, this work proposes a new algorithm, ADIFA  (\textbf{A}ttribute \textbf{D}ens\textbf{I}ty \textbf{F}unction \textbf{A}pproximation), which is later compared to three very known unsupervised anomaly detection algorithms that we transformed into one-class anomaly detection (see Section \ref{classifiers}). In the rest of this section, we focus on the ADIFA algorithm.\\

ADIFA is inspired by the Parzen-Window density estimation method \cite{parzen1962estimation}. Similar to Parzen-Window, ADIFA takes the approach of the non-parametric density estimation and assumes a probabilistic generative model for the observed data. However, unlike the Parzen-Window, ADIFA is a semi-supervised, meta-learning based algorithm. The algorithm learns multiple univariate models, each of which is responsible for approximating the density-function of a single related attribute, hence, permits per-attribute anomaly detection. Given a test XML document , the univariate density-function models compute a series of per-attribute normality scores. These scores are then combined via yet another univariate density-function model (i.e., the meta-classifier), which outputs the ADIFA prediction. We begin by discussing some preliminaries and then give a formal definition of ADIFA.

\begin{figure*}[ht]
	\begin{minipage}[l]{0.5\linewidth}
	\centering
		\includegraphics* [scale=0.6]{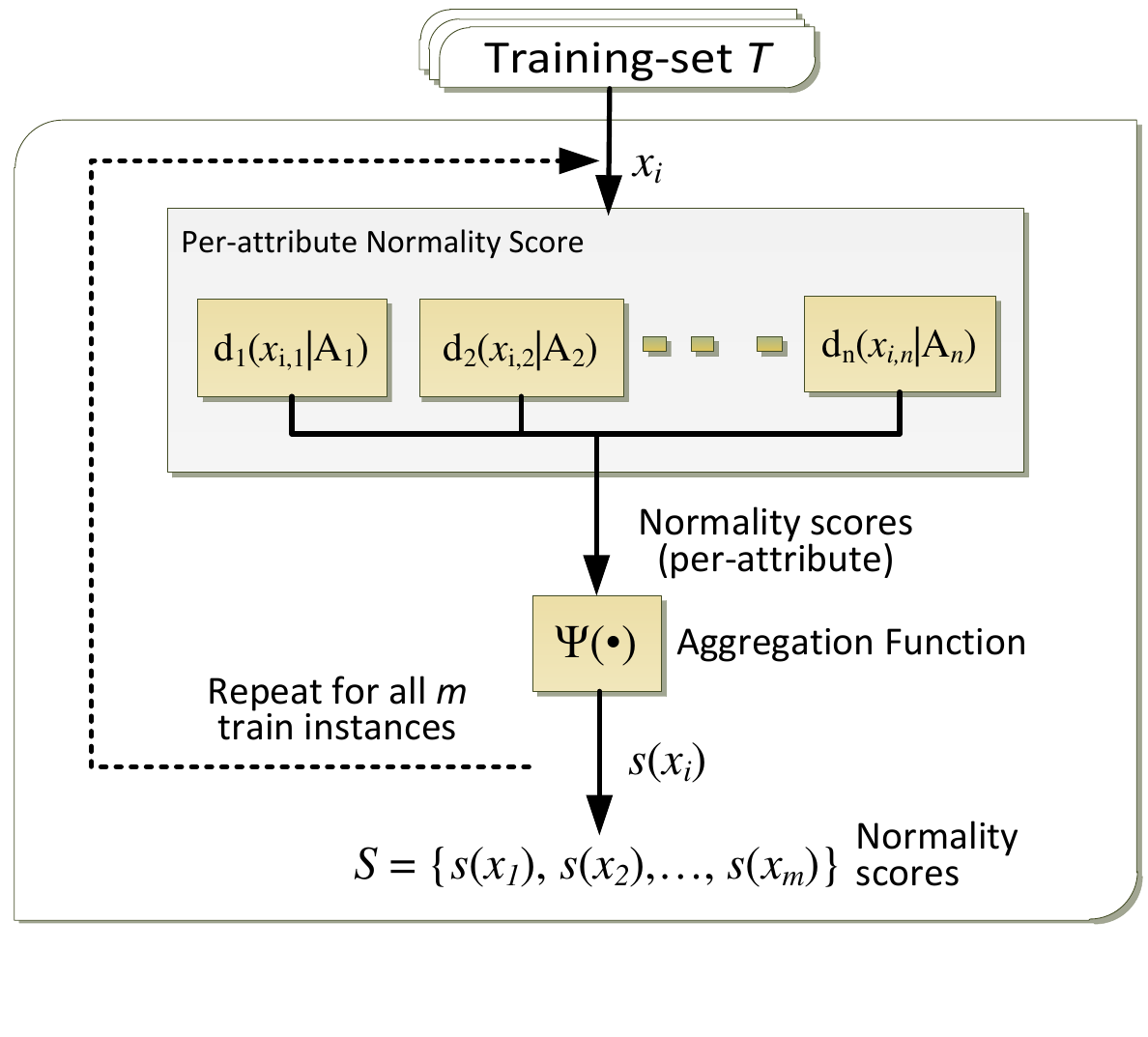}
		%\vspace{-3mm}
	\caption{ADIFA training process}\label{Fig:Training}
	%\vspace{-3mm}
\end{minipage}
\hspace{0.2cm}
\begin{minipage}[r]{0.5\linewidth}
	\centering
		\includegraphics* [scale=0.6]{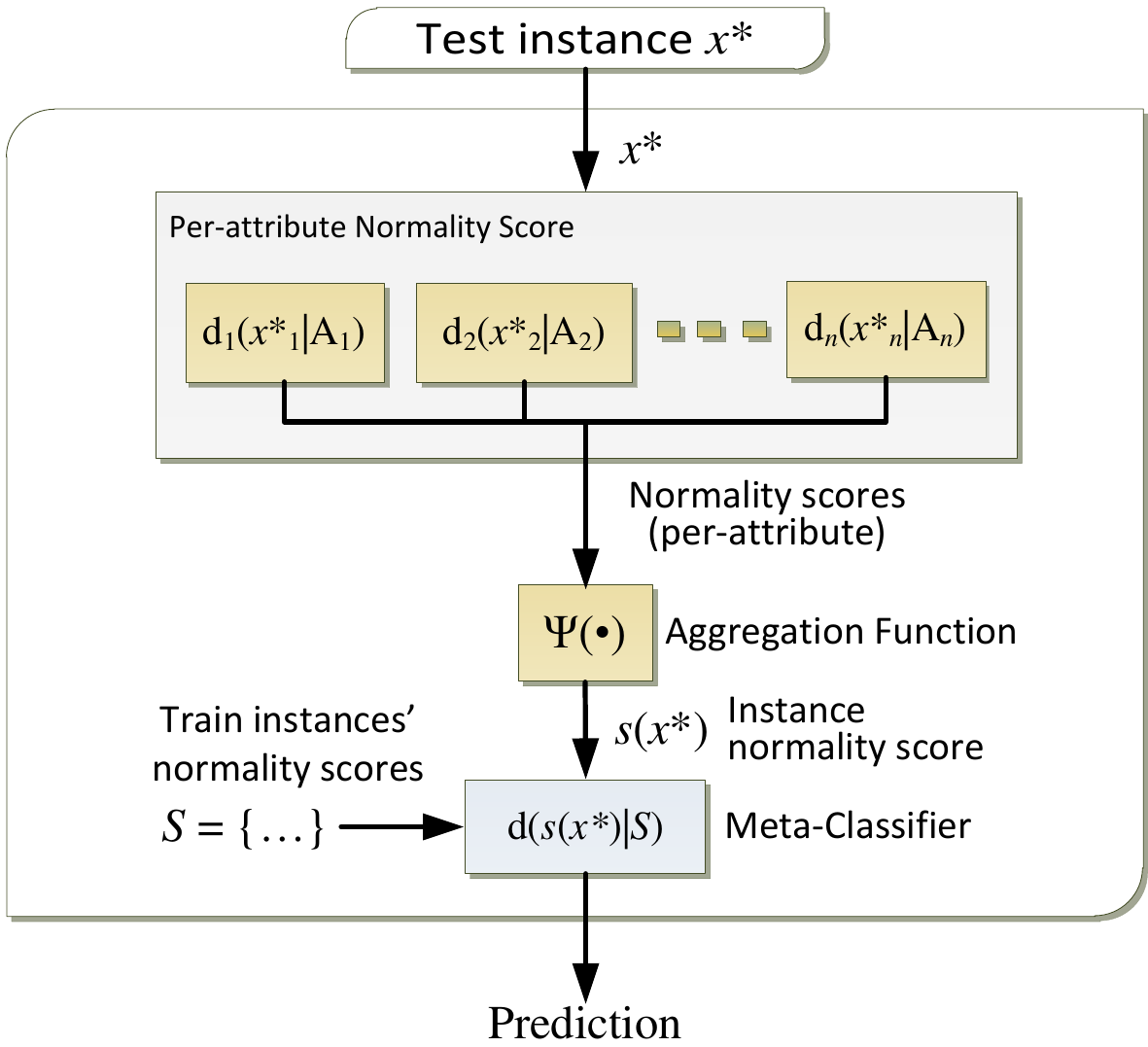}
		%\vspace{-3mm}
	\caption{ADIFA on-line classification process}\label{Fig:Classification}
	%\vspace{-3mm}
\end{minipage}
\end{figure*}

\subsubsection*{Preliminaries}
Let $T=\{x_1, x_2, \dots, x_m\}$ be a training set containing $m$ instances, drawn i.i.d. from $X^{\{normal\}}$, the group of \emph{normal} instances. We assume that the instances in $T$ well represent the \emph{normal} class.
%Since ADIFA learns multiple univariate models, one for each attribute of the dataset,
For our purpose, it is convenient to represent $T$ as a bag of $n$ independent features (attributes):
\[T=\{A_1,A_2, \dots , A_n\}\] where $A_j=\{a_{j,1},a{j,_2},\dots,a_{j,m}\}, j=1,\dots,n$ is the set of values of the $j^{th}$ attribute in $T$. We assume that $A_j$ is drawn i.i.d. from an unknown density function $d_j$. Let $x \in X$ be an instance to be classified. The ADIFA tasks are: (1) approximate the density functions $\{d_j\}_j={1,\dots,n}$; and (2) use these approximate models to calculate the likelihoods (or probability), that the instance $x$ was drawn from $X^{\{normal\}}$.\\
To safely use a set of univariate models for anomaly detection in multivariate vectors, as proposed in this algorithm, the next proposition should hold with high probability: $\forall{x_1, x_2}\in X$

\begin{center}
$D(x_1 | T)>D(x_2 | T) \Rightarrow$
\resizebox{1\linewidth}{!} {
$\Psi(d_1 (x_{1,1} | A_1 ),d_2 (x_{1,2} | A_2)\dots)>\Psi(d_1 (x_{2,1} | A_1),d_2 (x_{2,2} | A_2)\dots)$
}
\end{center}

Where $D(\cdot)$ is a multivariate anomaly detection model; $d(\cdot)$ is a per-attribute univariate anomaly detection model; and $\Psi(\cdot)$ is an aggregation function (e.g., arithmetic mean, geometric mean, and harmonic mean).

$D(x_1 | T)$ is the normality score of instance $x_1$, with respect to the train-set $T$, whereas $d_i (x_{1,i} | A_1 )$ is the normality score of the $i^{\text{\tiny th}}$ attribute of instance $x_1$.\\  
In other words, we assume that anomalous patterns can be effectively detected by aggregating attribute-wise normality scores.
Our experimental results show that in many domains, especially in computer and network security, this assumption holds. Notice that in order to cover all anomalous patterns, one should also address uncommon situations, in which examples have an anomalous \emph{combination} of normal values. This can be done, for example, by learning association-rules. 

\subsection{Training Process}
\label{ADIFA_Training}
First, a set of $n$ density-function models $D = \{d_1,d_2,\dots,d_n\}$ are learned. A model, $d_j:R \rightarrow[0,1]$ is responsible for approximating the density-function of the corresponding attribute, $Attr_j$, where $j=1,\dots, n$. 
To calculate the attribute-wise normality scores of a given test instance, $x^*=<x^*_1,\dots,x^*_n>$, we use the Gaussian radial-basis function (RBF), with a normalization factor $b_j=(2\pi\widehat{\sigma_j}^2)^{-\frac{1}{2}} $: 

\[\phi_j (x^*)= \frac{1}{m} \sum_{i=1}^{m} b_j * \rho _j (\left\|a_{i,j}-x^*_j\right\|), j=1,\dots,n \]  

Where $a_{i,j}$ is the $j$th attribute of train instance $i$, and $m$ is the cardinality of $A_j$. We chose the exponential-decay distance function for the RBF function:
\[\rho_j (\left\|a_{i,j}-x^*_j \right\|)=e^{-\tau_j(a_{i,j}-x^*_j)^2}\]  

The coefficient $\tau_j$ controls the similarity decay speed, which also controls the smoothness of the density function and therefore, its generalization power. The per-attribute, density approximation model, $d(\cdot)$ is defined as follows: 
 
\begin{equation}
\label{perFeatureAnomalyScore}
d_j (x^*_j |A_j)\equiv \phi_j (x^*)=\frac{1}{m} \sum_{i=1}^{m}\frac{1}{\sqrt{2\pi \widehat{\sigma_j}^2}}*e^{-\tau_j (a_{i,j}-x^*_j)^2}
\end{equation}

To compute the instance-wise normality score, $s(x)$, we use an aggregate function $\Psi(\cdot)$ on the weighted per-attribute normality scores as follows:%, as shown in equation \ref{InstanceNormalityScore}. 

\begin{equation} 
\label{InstanceNormalityScore}
\resizebox{0.91\linewidth}{!} {
s(x)=$\Psi \left(\alpha_1 \ast d_1 (x_1 | A_1), \alpha_2 \ast d_2(x_2 | A_2),\dots, \alpha_n \ast d_n (x_n | A_n)\right)$
}
\end{equation}

Where the weights, $\alpha_{1,\dots, n}$, reflect the complexity of learning the corresponding approximation models. The weights can be obtained by using methods such as entropy-based weighting 

\begin{equation} 
\label{AttributeWeighting}
\alpha_j=1-\frac{H(A_j)}{\Sigma_{i=1}^nH(A_i)}
\end{equation}

or by using techniques such as\cite{DBLP:conf/approx/GottliebK10} to efficiently estimate the per-attribute doubling dimension.

The notation $H(A_j)$ in Equation \ref{AttributeWeighting} specifies the entropy of attribute $A_j$. Note that the lower the attribute's entropy is, the easier it is to accurately approximate (i.e. to learn) its density function. Since 
$\frac{H(A_j)}{\Sigma_{i=1}^nH(A_i)} \in[0,1]$, 
the weights, $\alpha_j$, calculated by this technique are also in the range of $[0,1]$. \\

Lastly, the training applies Equation \ref{InstanceNormalityScore} to each of the $m$ instances to compute the normality scores of the training instances. The training employs a leave-one-out scheme i.e. each training instance is used only once as a test instance, while the rest $m-1$ instances are used as the training set $T$. Let $S$ be the normality scores group of the training instances, i.e., $S = \{s(x_1),s(x_2),\dots,s(x_m)\}$.

\subsection{Classification with ADIFA}
The classification of a test sample is done in two steps. First, the test instance's normality score, $s(x^*)$, is calculated using the technique presented in the previous section. Then, the likelihood of obtaining such a normality score, according to the density function, $p(x)$, of the normal instances, is computed.

Assuming that the normality scores in $S$ are drawn i.i.d. from $p(x)$, the density function $p(x)$ can be approximated by a single univariate model. This is done using the same technique presented in Equation \ref{perFeatureAnomalyScore}.

In order to calculate the likelihood of obtaining $s(x^*)$, the algorithm computes how anomalous $s(x^*)$ is with respect to $S$. This is done by approximating the density function $p(x)$ by using an additional univariate density approximate, with the parameters $s(x^*)$ and $S$:  

\begin{equation}
\label{InstanceAnomalityScore} 
%l(s(x^*)) =  
d(s(x^*) | S) = \frac{1}{m} \sum_{i=1}^m \frac{1}{\sqrt{2\pi \widehat{\sigma_S}^2}}*e^{-\tau_S (S(x_i)-S(x^*))^2}  
\end{equation}
 
The value obtained by Equation \ref{InstanceAnomalityScore} is the output of ADIFA. In case a classification is needed, this value can be thresholded by a user defined value $0<C\leq 1$. In this case, ADIFA predicts \emph{anomaly} if $d(s(x^*) | S)<C$ and \emph{normal} otherwise.

\subsection{Anomaly Localization Strategy}
Identifying the location of an anomaly within the transaction document is a task that the system or service security officer has to deal with when a transaction is suspected of being anomalous. Locating the anomaly can be very important, particularly when the anomaly source may indicate an ongoing attack. Since the location of an attack-related anomaly is known, some mitigation measures can be employed. It is also important for the post-attack analysis, since the location information may provide the forensic expert with helpful facts regarding the attacker's sources, attack methods and propagation.  Although such localization will not provide a complete operational meaning, such as the anomaly semantic, it provides valuable information towards this goal.

Once an anomaly is detected, the anomaly localization is straightforward. Since it is a multi-univariate classifier, the ADIFA classifier has full knowledge regarding the normality score of each of the test instance dimensions. When an anomalous transaction is detected, the ADIFA classifier identifies the features with the lowest likelihoods. Additionally, the classifier can produce a list of all features, ordered by their likelihood values. 

\section{Methods}
\label{methods}
In this Section we specify the methods for evaluating the algorithms proposed in this paper. First, in Section \ref{classifiers}, we describe the classifiers we used. Then in Section \ref{metrics}, the performance metrics are detailed. Next, in Section \ref{datasets}, we present the datasets used in our experiments. Lastly, in section \ref{AnomalyInjector} we present an application for injecting real attacks in XML transactions.

\subsection{Classifiers}
\label{classifiers}
We made use of six classifiers trained by four anomaly detection (one-class) algorithms: OC-GDE, OC-PGA, 1-SVM \cite{Schlkopf99estimatingthe}, and ADIFA\footnote{The implementation of the mentioned algorithms can be downloaded from: \url{http://sourceforge.net/projects/xml-ad/files/AD.rar/download}}. We selected these classifiers, as they represent the prominent branches of one-class algorithms: nearest neighbor (OC-PGA), density estimation (OC-GDE) and boundary (1-SVM). The first three algorithms are our own adaptations of two well-known unsupervised algorithms to one-class learning. Table \ref{tab:ClassifiersSetup} specifies the setup parameters of all seven classifiers used for the  experiments. Note that for ADIFA classifiers, we used equation \ref{AttributeWeighting} to compute the individual features weights. We used a classification threshold, $C=0.5$ for all of the classifiers, i.e., instance $x\in$ \emph{normal} if $p(x) \leq 0.5$, where $p(x) \in [0,1]$ is the classifier output.

\subsubsection*{One-Class Peer Group Analysis}
The One-Class Peer Group Analysis (OC-PGA), is an adaptation of the unsupervised Peer-Group-Analysis method (PGA), proposed by Eskin et al. \cite{Eskin02ageometric} for the one-class learning domain. The algorithm identifies anomalies as points in low-density regions of the feature space. An anomaly score is computed at point $x$ as a function of the distances from $x$ to its $k$ nearest neighbors. Although PGA is actually a ranking technique applied to a clustering problem, we implemented it as a one-class classifier. Given a training set $S$, a test point $x$ is classified as follows. For each $x_i\in S$, we pre-compute the distance from $x_i$ to its nearest neighbor in $S$, and denote it by: $d_{nn}(x_i,S\setminus\{x_i\})$.
To classify $x$, the distance to its nearest neighbor in $S$, $d_{nn}(x,S)$ is computed. The test point $x$ is classified as an anomaly if $d_{nn}(x,S)$ is within a percentile $\alpha$ or higher among $\{d_{nn}(x_i, S\setminus\{x_i\} )\}_{x_i \in S}$ ; otherwise, it is classified as \emph{normal}. 

\subsubsection*{One-Class Global Density Estimation}
The Global Density Estimation (GDE), proposed in \cite{Knorr97aunified}, is also an unsupervised density-estimation technique, which uses the nearest-neighbor technique. Given a training set $S$ and a real value $r$, one computes the anomaly score of test point $x$ by comparing the number of training points falling within the $r$-ball $B_r (x)$ about $x$ to the average of $\left|B_r(x_i)\cap S\right|$ over all $x_i \in S$. We set $r$ to be twice the set average of $d_{nn}=(x_i,S\setminus\{x_i\})$  to ensure that the average number of neighbors is at least one. 
To adapt the GDE into the one-class domain (OC-GDE), we used the following heuristic function to threshold the anomaly scores, since it achieved a low classification error on the data:
\emph{$x$ is labeled as \emph{normal} if $e(-(n_r(x)-\overline{N_r})/{\sigma_r})>1/2$} where $n_r(x)$ is the number of neighbors of $x$ in $B_r (x)\cap S$ (points in $S$ at a distance no higher than $r$ from $x$), $\overline{N_r}$ and $\sigma_r$  are the average and standard deviation of  $\{n_r(x_i)\}_{x_i \in S} $, respectively.

\begin{table}[htbp]
	\centering

		\begin{tabular}{@{}l@{\hspace{1mm}}c@{\hspace{1mm}}c@{\hspace{1mm}}l@{}}
		\hline
		Algorithm & Classifier & Acronym & \multicolumn{1}{c}{Parameters} \\
		\hline
		ADIFA 	& ADIFA-AM 	& AD-AM & $\Psi$ = Arithmetic Mean  \\
		ADIFA 	&	ADIFA-HM 	& AD-HM & $\Psi$ = Harmonic Mean  \\
		ADIFA 	& ADIFA-GM 	& AD-GM & $\Psi$  = Geometric Mean \\
		OC-GDE 	& OC-GDE 		& GDE 	& $-$ \\
		OC-PGA 	& OC-PGA 		& PGA 	& $k=1$ (1-nearest neighbor), $\alpha=0.1$  \\
		%OC-LOF	& OC-LOF & OC-LOF  	& $-$ \\
		1-SVM 	& 1-SVM 		& SVM 	& Kernel = RBF (Gaussian), $\nu=0.05$ \\
		\hline
		\end{tabular}
	\caption{Classifier's setup parameters. The parameters shown are only those that are non-default.}
	%\vspace{-3mm}
	\label{tab:ClassifiersSetup}
\end{table}

\subsection{Performance Metrics}
\label{metrics}
The main measure used to evaluate the classification performance of XML-AD was the area under the receiver operating characteristic (ROC) curve, which is a graphical plot of the specificity vs. 1-sensitivity for a classifier system since its discrimination threshold is varied. The ROC can also when represented equivalently by plotting the fraction of true positives (TPR = true positive rate) vs. the fraction of false positives (FPR = false positive rate). ROC analysis provides tools for selecting possible optimal models and discarding suboptimal ones independently from (and prior to specifying) the cost context or the class distribution. ROC analysis is strongly related to cost/benefit analysis in diagnostic decision-making. Widely used for many decades in medicine, radiology, psychology, and other areas, it has been introduced rather recently into machine-learning and data mining.\\ 
In order to estimate the area under ROC (AUC), a 5x2 cross-validation procedure was performed \cite{Dietterich98}. In each of the cross-validation iterations, the training set was randomly partitioned into two disjoint instance subsets. In the first fold, the first subset was utilized as the training set, while the second subset was utilized as the testing set. In the second fold, the role of the two subsets was reversed. This process was repeated five times. The same cross-validation folds were implemented for all algorithms in all experiments. 

The one-tailed paired $t$-test with a confidence level of 95\% verified whether the differences in AUC between tested classifiers were statistically significant. To conclude which classifier performed best over multiple datasets, we followed the procedure proposed in \cite{Demsar06}. We first used the adjusted Friedman test to reject the null hypothesis, followed by the Bonferroni-Dunn test to examine whether a specific classifier produces significantly better AUC results than the reference method.

\subsection{XML transactions Collections}
\label{datasets}
To evaluate the XML-AD framework we used three transaction collections.The first two, Insurance and Inventory, are collections of XML transactions of real information system. They contain only transactions related to the \emph{normal} class. i.e., they do not contain any attacks or anomalies of interest. The third collection, ARP-D, contains XML transactions derived from ARP packets, which were captured on the local network at the BGU university (2011). This collection contains ARP transactions related to an actual ARP attack. 

\begin{figure*}[t!]
	\centering
	\resizebox{1\linewidth}{!} {
		\includegraphics{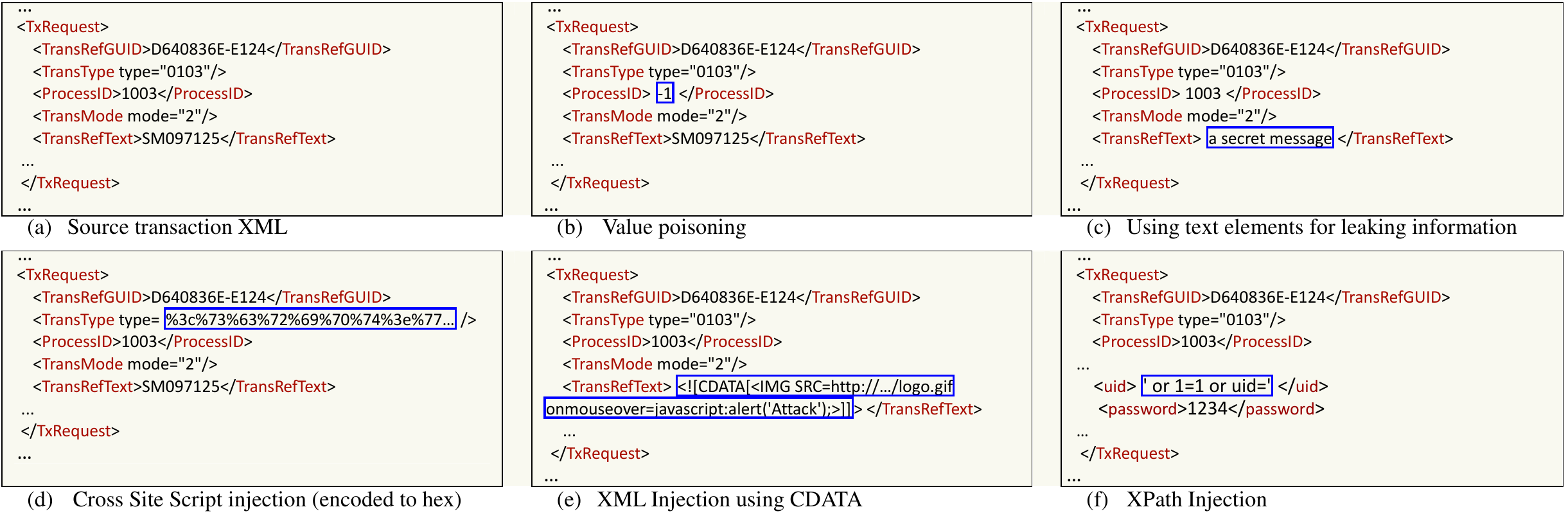}
		}
	\caption{Five types of XML attacks, embedded in target XML file, shown in (a), using the AI module}
	\label{fig:Synthesizer}
\end{figure*}

\subsubsection*{Insurance \& Inventory}
\label{XML-Transactions}
The Insurance collection contains 3,340 insurance transactions, taken from a real insurance information system that follows the ACORD standard \cite{acord} – all of which are labeled as \emph{normal}. Since the insurance XML transactions contain many private data items, a pre-process of anonymization was employed to ensure the privacy of the insurees. 
The second collection, Inventory, is comprised of transactions that were extracted from a logistic information system which mainly consists of supply data. The collection contains 4,000 transaction, all labeled as \emph{normal}. \\
Both XML transactions collections were put through the processes of feature extraction and dataset-flattening, described in Sections \ref{FeatureExtraction} and \ref{datasetGeneration}. These processes extracted 1,021 features from each transaction in the Insurance collection, and 285 features from each transaction in the Inventory collection, regardless of the original transaction size or the number of elements.

\subsubsection*{ARP-D}
\label{ARP-D}
The ARP abbreviation stands for “Address Resolution Protocol” \cite{plummer1982rfc}. The ARP-D dataset contains actual ARP spoofing attacks directed against the computer network of Ben-Gurion University. During an ARP attack, the attacker temporarily steals the IP addresses of its victims and as a result all of their traffic was redirected to the attacker without the knowledge or consent of the victims.
The collection is constructed from 9,039 ARP packets, captured, and converted to XML using the Wireshark tool \cite{Wireshrk} at the local-network, for the duration of 10 minutes. The ARP spoofing attack began 5 minutes after the Wireshark tool had began sniffing the network traffic and lasted for 5 additional minutes. During the recording time period, there were 173 active computers on the local network, 27 of which were attacked. 
The XML files in this collection were put through a process of feature extraction and dataset-flattening, similar to the process that was applied to the Insurance and Inventory collections. This process extracted 24 attributes from each ARP XML transaction.
In order to examine different feature aggregation methods, two distinct datasets ARP-D$_1$ and ARP-D$_2$, which differ by their feature aggregation properties, were constructed.

\subsection{Anomalous XML Transactions}
\label{AnomalyInjector}

To properly evaluate XML-AD, it was necessary to experiment with both \emph{normal} and \emph{anomalous} XML transaction documents from real systems. However, in many real-life systems, as is the case with the Insurance and Inventory systems, XML documents containing anomalies of interest (i.e., anomalies that can potentially harm the system), are extremely rare, to the point where entire data collections can be safely presumed to be \emph{normal}. 

To overcome this problem, we implemented an XML anomaly injection (AI) module, which embeds instances of \textbf{real XML attacks} into target XML transactions, which are otherwise \emph{normal} XML transactions. 
The AI module implements five different attack classes: Value poisoning, Cross-site scripting (XSS), an XML injection using CDATA, an XPAth injection, and Data leakage in text elements. 

The Value poisoning attack is performed by changing a single numerical element in the target XML transaction to a random value. The XSS and the XPath attacks are represented by strings (either as clear or encoded text) and are embedded into a randomly chosen textual XML element. The CDATA XML injection attack embeds four different malicious encoded java scripts into CDATA elements. A CDATA XML attacks is injected between two randomly selected adjacent XML elements of the target XML transaction. Lastly, the data-leakage attack is performed by replacing the content of a textual XML element with randomly selected sentences from the known children book: \emph{The wonderful Wizard of Oz}. The mentioned XML attack are demonstrated in Figure \ref{fig:Synthesizer}.

The AI module accepts two parameters: a target XML transaction and an anomaly index, which is the ratio between the number of the anomalies to inject and the number of XML elements in the target XML transaction.

When constructing an anomalous XML transaction, the AI module embeds one or more instances of the mentioned attack classes into the target XML transaction at random. In order to avoid being ruled out by the XML parser, the AI module embeds the attacks only where it is allowed according to the XSD schema, .i.e., XPath injections are only embedded into textual XML elements, Value poisoning is only applied to numeric elements, etc. 
 
Notice that the above mentioned attacks are bound to introduce anomalies into the target XML transactions because they represent very rare events, with respect to the content of the \emph{normal} transactions. These anomalies may or may not maliciously affect the consumer information system. Either way, such transactions should be detected as anomalous due to the reasons mentioned in Section \ref{problemStatement}. Finally, we would like to stress that in order to train the anomaly detection classifiers, only \emph{normal}-labeled XML documents were used. The anomalous XML transactions were only used for testing.

\section{Evaluation}
\label{evaluation}
The evaluation of the XML-AD framework (Fig. \ref{fig:ADIFA_Process}) focused on (a) determine whether the proposed approach can effectively detect anomalies in XML documents and (b) which machine-learning approach is best suited for that task. Regarding the ADIFA learning algorithm, we examined which of the proposed aggregation functions produces the best classification performance. Additionally, we examined the anomaly detection capabilities of ADIFA in other domains and compared it to similar algorithms. To examine the abovementioned aspects, we implemented an evaluation framework according to the scheme shown in Figure \ref{fig:ADIFA_Process} and the classifiers in Section \ref{classifiers}. The following sections describe the experiments and results.

\subsection{Anomaly Detection Framework}
In this section we present our evaluation of the XML-AD framework in detecting anomalous XML transactions. 
First, we investigate the performance of the available one-class classifiers when applied to our XML transactions datasets. Next, we examine the sensitivity of the classifiers to the level of abnormality, within the XML attacks. Finally, we show the learning curves of our one-class classifiers.

\subsubsection{XML Anomaly Detection}
Our first experiment evaluates the dataset flattening and the anomaly detection aspects of the XML-AD framework. First, we examine whether the dataset flattening process damages the ability to detect XML attacks, due to the lossy nature of the flattening process. Next, we compared the two classification approaches, i.e., the multivariate and multi-univariate approaches, to determine which one, if any, achieves a better detection of XML anomalies. 

\begin{figure}[b]
	\centering
		\resizebox{1\linewidth}{!} {
		\includegraphics{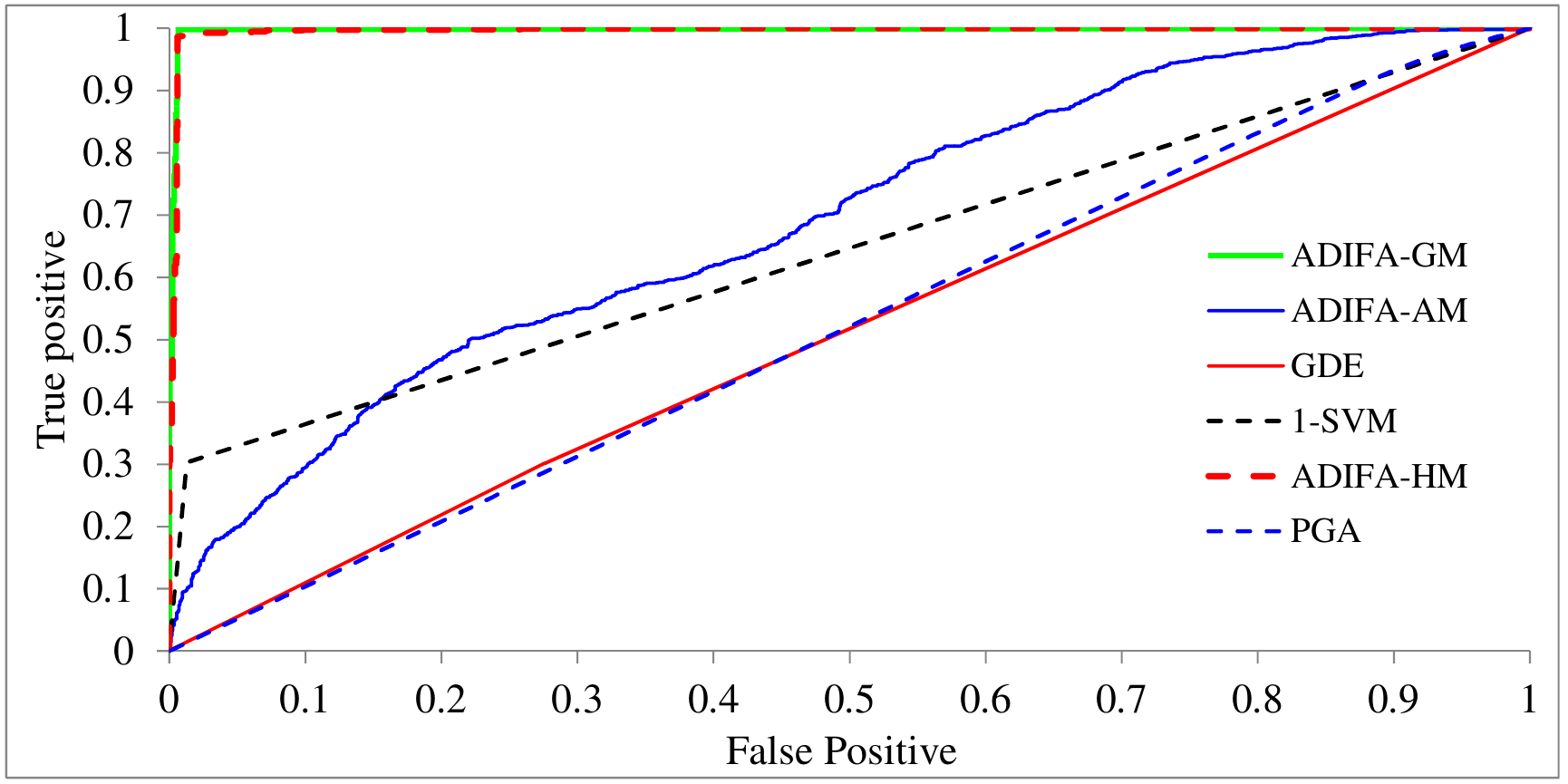}
		}
	\caption{ROC plot for \emph{Insurance} dataset with 5\% anomalous XML elements in the abnormal transactions}
	%\vspace{-3mm}
	\label{fig:InventoryROC}
\end{figure}

The 1-SVM, OC-GDE and OC-GDE classifiers belong to the multivariate approach, whereas the ADIFA-GM, ADIFA,HM and ADIFA-AM classifiers are multi-univariate.
The experiment results are presented in Table \ref{tab:ResultTableForXMLTransaction}. The classifiers` ROC curves, which are depicted in Figure \ref{fig:InventoryROC}, show that it is possible to obtain a detection rate of 89\%, with only 0.2\% false positive rate.

\begin{table}[t]
	\centering
	%\scriptsize
		\begin{tabular}{@{}p{40pt}@{\hspace{-2mm}}p{15mm}@{\hspace{0.3mm}} @{\hspace{0.3mm}}c@{\hspace{0.3mm}} @{\hspace{0.3mm}}c@{\hspace{0.3mm}} @{\hspace{0.3mm}}c@{\hspace{0.3mm}} @{\hspace{0.3mm}}c@{\hspace{0.3mm}} @{\hspace{0.9mm}}c@{\hspace{0.3mm}} @{\hspace{0.3mm}}c@{}}
		\hline
		
		 	\multirow{2}{*}{Datasets}	& \multicolumn{1}{c}{Anomaly} & \multicolumn{6}{c}{Classifiers} \\  \cline{3-8}
			
		  & \multicolumn{1}{@{\hspace{0.3mm}}c}{index} & AD-AM & AD-GM & AD-HM & GDE & 1-SVM & PGA \\
		\hline
		\multirow{3}{*}{Insurance} & \multicolumn{1}{c}{1\%} & 0.570 & 0.951 & 0.944 & 0.498 & 0.501 & 0.524 \\
															 & \multicolumn{1}{c}{5\%} & 0.692 & 0.996 & 0.995 & 0.574 & 0.498 & 0.532 \\
															 & \multicolumn{1}{c}{10\%} & 0.818 & 0.998 & 0.999 & 0.589 & 0.504 & 0.539 \\
		\hline
		%\vspace{1mm}
			\multicolumn{2}{r}{Average} & 0.693 & 0.981 & 0.979 & 0.554 & 0.501 & 0.532 \\
		\hline
		\multirow{3}{*}{Inventory} 	& \multicolumn{1}{c}{1\%} & 0.590 & 0.887 & 0.884 & 0.500 & 0.537 & 0.527 \\
																& \multicolumn{1}{c}{5\%} & 0.683 & 0.956 & 0.954 & 0.501 & 0.562 & 0.642 \\
																& \multicolumn{1}{c}{10\%} & 0.757 & 0.984 & 0.982 & 0.501 & 0.575 & 0.744 \\
		\hline
		%\vspace{1mm}
			\multicolumn{2}{r}{Average}  & 0.677 & 0.943 & 0.940 & 0.501 & 0.558 & 0.638 \\
		\hline
		ARP-D$_1$ & \multicolumn{1}{c}{n/a} & 0.873 & 0.871 & 0.799 & 0.634 & 0.643 & 0.929 \\
		ARP-D$_2$ & \multicolumn{1}{c}{n/a} & 0.900 & 0.910 & 0.926 & 0.635 & 0.635 & 0.936 \\
		\hline
		%\vspace{1mm}
			\multicolumn{2}{r}{Average} & 0.980 & 0.969 & 0.949 & 0.639 & 0.634 & 0.933 \\
		\hline
		Total & & 0.759 & \myem{0.964} & 0.957 &  \mybf{0.555} &  \mybf{0.556} & 0.672 \\

		\hline		\end{tabular}
	\caption{Average AUC result for the XML transactions datasets}
	\label{tab:ResultTableForXMLTransaction}
	%\vspace{-3mm}
\end{table}

These results show that for each tested dataset, there was at least one classifier that successfully distinguished between the anomalous XML transactions and the normal XML transactions. This leads to the conclusion that the lossy flattening process preserves the information needed for detecting XML anomalies. 

Examining the classifiers performance, we notice that ADIFA-GM achieves the highest average AUC among all the tested classifiers. The second best result was achieved by ADIFA-HM (AD-HM). Both results are substantially better than the other four classifiers. The GDE and 1-SVM classifiers performed the worst, with an average AUC close to 0.5, a value that represents a random classifier. By further analyzing the results, it is clear that the choice of aggregation functions plays a crucial part in ADIFA. In datasets where the dimensionality is relatively low (ARP-D$_1$ and ARP-D$_2$), the arithmetic-mean yields better results, whereas in higher-dimensionality datasets, such as with Insurance and Inventory, ADIFA performs better when it used with the geometric-mean aggregation function. Interestingly, the results show that all three multivariate classifiers, i.e., GDE, PGA, and 1-SVM, performed very poorly on Insurance and Inventory, the datasets which contain genuine XML transactions. These results support our hypothesis regarding the uselessness of multivariate anomaly detection models (models that use multivariate distance function) for detecting anomalies in XML transactions.

\subsubsection{Multivariate Vs. Multi-Univariate Classifiers}
%We examine the detection performance as a function of the fraction of anomalous elements in the XML transactionassifiers}
The previous experiment showed that the ADIFA classifiers performed substantially better than the other multivariate classifiers, especially when applied to high dimensionality datasets. To explain these results, we examine the following hypotheses:

\begin{itemize}
\item $h^1$ - multi-univariate classifiers are inherently more accurate than multivariate classifiers, when dealing with anomaly detection in high-dimensional datasets. 
\item $h^2$ - the classification performance depends on the data modeling method, i.e., compared to the other classifiers, the ADIFA algorithm better modeled the information, which resides within the tested datasets, and hence, it produced a superior anomaly detection classifier.
\end{itemize}

To examine the hypotheses, we conducted an experiment in which we transformed the multivariate classifiers into multi-univariate systems by training each multivariate classifier with a single dimension and then aggregating their results using the geometric-mean. Finally, we compared the new univariate systems with ADIFA-GM, which also uses the geometric-averaging aggregation. We denote the new systems as MU$_{SVM}$, MU$_{GDE}$ and MU$_{PGA}$, where the subscript indicates the base classification algorithm and MU stands for multi-univariate.

In this experiment we also measured the time it takes to classify the datasets. Classification time (CT) is an important feature of the anomaly detector, since it dictates the rate in which transactions can be classified in real-time. To avoid being a bottleneck for the XML transaction consuming process, the classification time should be as low as possible.

\begin{table}[h]
		\resizebox{1\linewidth}{!} {
	\centering
	
		\begin{tabular}{@{}p{30pt}@{}p{10pt}@{}c@{\hspace{0.7mm}}c@{\hspace{0.7mm}}|c@{\hspace{1.5mm}}c@{\hspace{1mm}}@{\hspace{1mm}}|c@{\hspace{0.7mm}}c@{\hspace{0.7mm}}|c@{\hspace{2mm}}c@{}}
		\hline
	\multirow{3}{*}{Datasets}	
	& \multicolumn{1}{c}{Anomaly} & \multicolumn{8}{c}{Classifiers} \\ \cline{3-10} 
	& \multicolumn{1}{@{\hspace{0.3mm}}c}{index} & \multicolumn{2}{c|}{AD-GM}  & \multicolumn{2}{c|}{MU$_{SVM}$} & \multicolumn{2}{c|}{MU$_{GDE}$} & \multicolumn{2}{c}{MU$_{PGA}$}  \\
	
	&  	& \scriptsize AUC & \scriptsize CT & \scriptsize AUC & \scriptsize CT	& \scriptsize AUC & \scriptsize CT	& \scriptsize AUC & \scriptsize CT \\
	\hline

\multirow{3}{*}{Insurance} 	
							& \multicolumn{1}{c}{1\%}		& 0.951 & 0.8 & 0.477 & 2.5 & 0.506 & 32.1 & 0.861	& 41.3 \\
							& \multicolumn{1}{c}{5\%} 	& 0.996 & 0.8 & 0.497 & 3.0 & 0.506 & 34.9 & 0.905  & 69.9 \\
							& \multicolumn{1}{c}{10\%}	& 0.998 & 0.7 & 0.521 & 2.3 & 0.506 & 29.0 & 0.905 	& 33.7 \\
							\hline
							\multicolumn{2}{r}{Average} & 0.981 & 0.8 & 0.498 & 2.6 & 0.506 & 31.9 & 0.890 & 47.6 \\
							\hline
\multirow{3}{*}{Inventory} 	
							& \multicolumn{1}{c}{1\%} 	& 0.888 & 0.4 & 0.474 & 1.4 & 0.530 & 15.0 & 0.820 & 49.3 \\
							& \multicolumn{1}{c}{5\%} 	& 0.960 & 0.4 & 0.486 & 1.6 & 0.530 & 18.7 & 0.885 & 84.1 \\
							& \multicolumn{1}{c}{10\%} 	& 0.984 & 0.4 & 0.496 & 1.4 & 0.533 & 16.8 & 0.910 & 33.8 \\
							\hline
							\multicolumn{2}{r}{Average} & 0.945 & 0.4 & 0.485 & 1.5 & 0.531 & 16.9 & 0.872 & 56.7 \\
							\hline
ARP-D$_1$ & \multicolumn{1}{c}{n/a} & 0.966 & 0.2 & 0.758 & 1.8 & 0.840 & 2.2 & 0.959 & 2.1 \\
ARP-D$_2$ & \multicolumn{1}{c}{n/a} & 0.973 & 0.2 & 0.610 & 1.6 & 0.873 & 2.6 & 0.963 & 2.5 \\
\hline
\multicolumn{2}{r}{Average} 		& 0.969 & 0.2 & 0.684 & 1.7 & 0.857 & 2.4 & 0.961 & 2.3 \\
\hline
Total 							& 	& \myem{0.964} & \myem{0.5} & \mybf{0.540} & 2.0 & 0.603 & 18.9 & 0.901 & \mybf{39.6} \\
\hline 
\end{tabular}
}
\caption{Average classification AUC and classification time (sec.) for the multi-univariate systems on the XML transactions datasets}

\label{tab:univariateResults}

\end{table}

The evaluation results, shown in table \ref{tab:univariateResults}, indicate that, on average, transforming a multi-variate classifier into multi-univariate system increases its classification effectiveness. However, the results are not consistent; the classification performance improvements were +34.2\% , +8.6\%  and -2.9\% for the MU$_{PGA}$, MU$_{GDE}$ and the MU$_{SVM}$ ,respectively, compared to their multivariate versions. The average performance boost, however, was not enough to outperform the ADIFA algorithm, which, as in the previous experiment, achieved the highest classification efficiency (AUC$=0.964$) among the tested classifiers. \\
Inspecting the classification time, we notice that the classification systems produce a wide range of results. The average classification time of the slowest classification system, i.e., the MU$_{PGA}$, was about 40 seconds, which is about two orders of magnitude higher than the minimal classification time, achieved by the ADIFA algorithm.
Lastly, the presented results validate both $h^1$ and $h^2$.

\subsubsection{ Classifier Performance Vs. Anomaly Index}
In the following experiment, we tested the classifiers' responsiveness to XML anomalies. This experiment had two goals: 1) to learn how responsive each classifier was to different levels of transaction anomalies and 2) to find interesting trends, such as, which classifier was more effective for detecting anomalies in datasets containing a small percentage of anomalies and which was preferable for higher rates of anomalies. Such a scenario is indeed possible since the classifier improvement rate,  as a function of the anomaly index, can differ from one classifier to the other. If such trend were to be found, an ensemble of classifiers would probably offer the best anomaly detection solution. To accomplish this experiment our six classifiers were applied to ten variations of the Insurance and Inventory datasets, each with a distinct anomaly index (1 to 10). 

\begin{figure}[h]
	\centering
		\resizebox{1\linewidth}{!} {
		\includegraphics{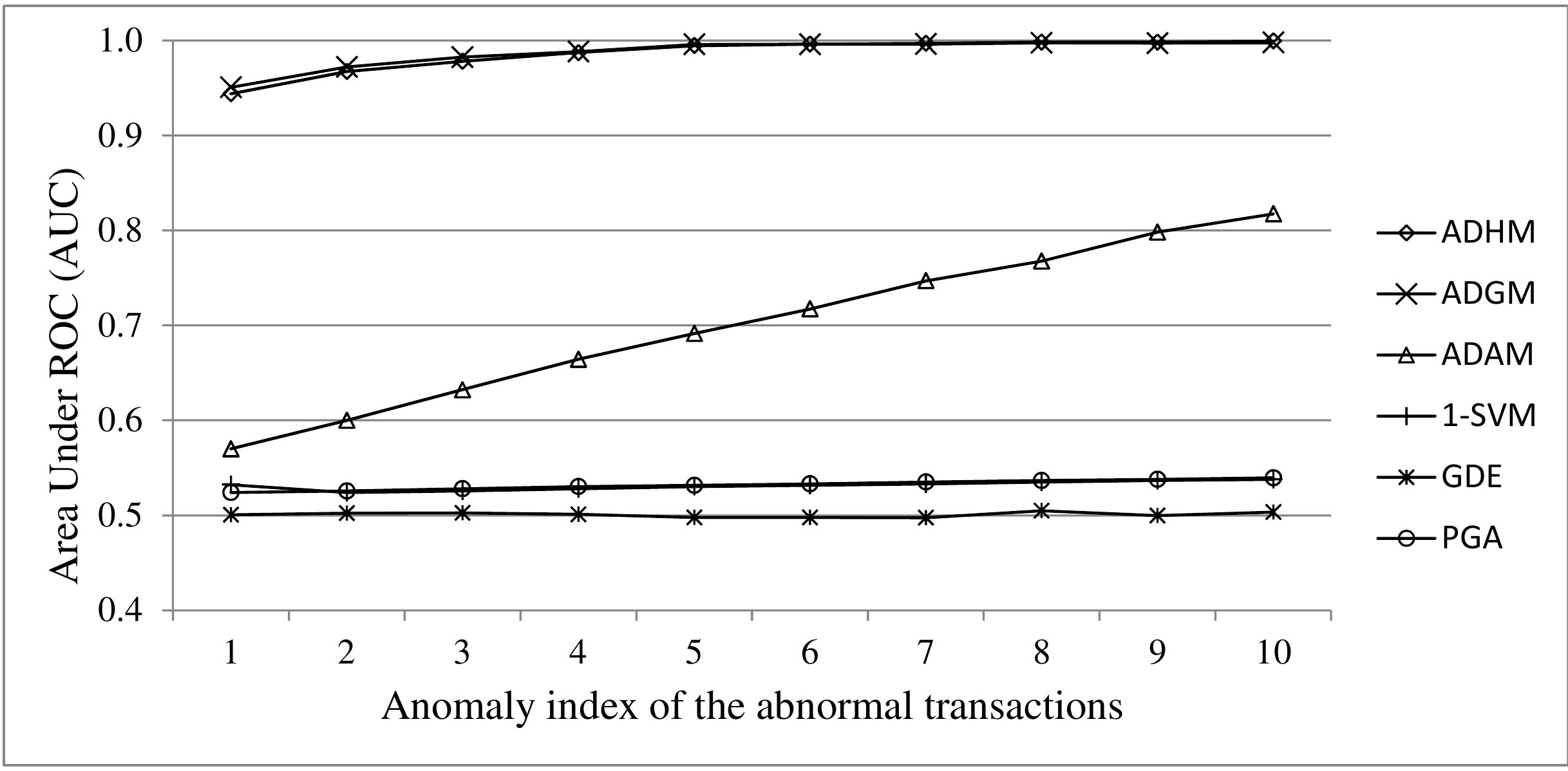}
		}
	\caption{AUC vs. XML anomaly index (\emph{Insurance dataset})}
	\label{fig:Insurance}
\end{figure}

\begin{figure}[h]
	\centering
		\resizebox{1\linewidth}{!} {
		\includegraphics{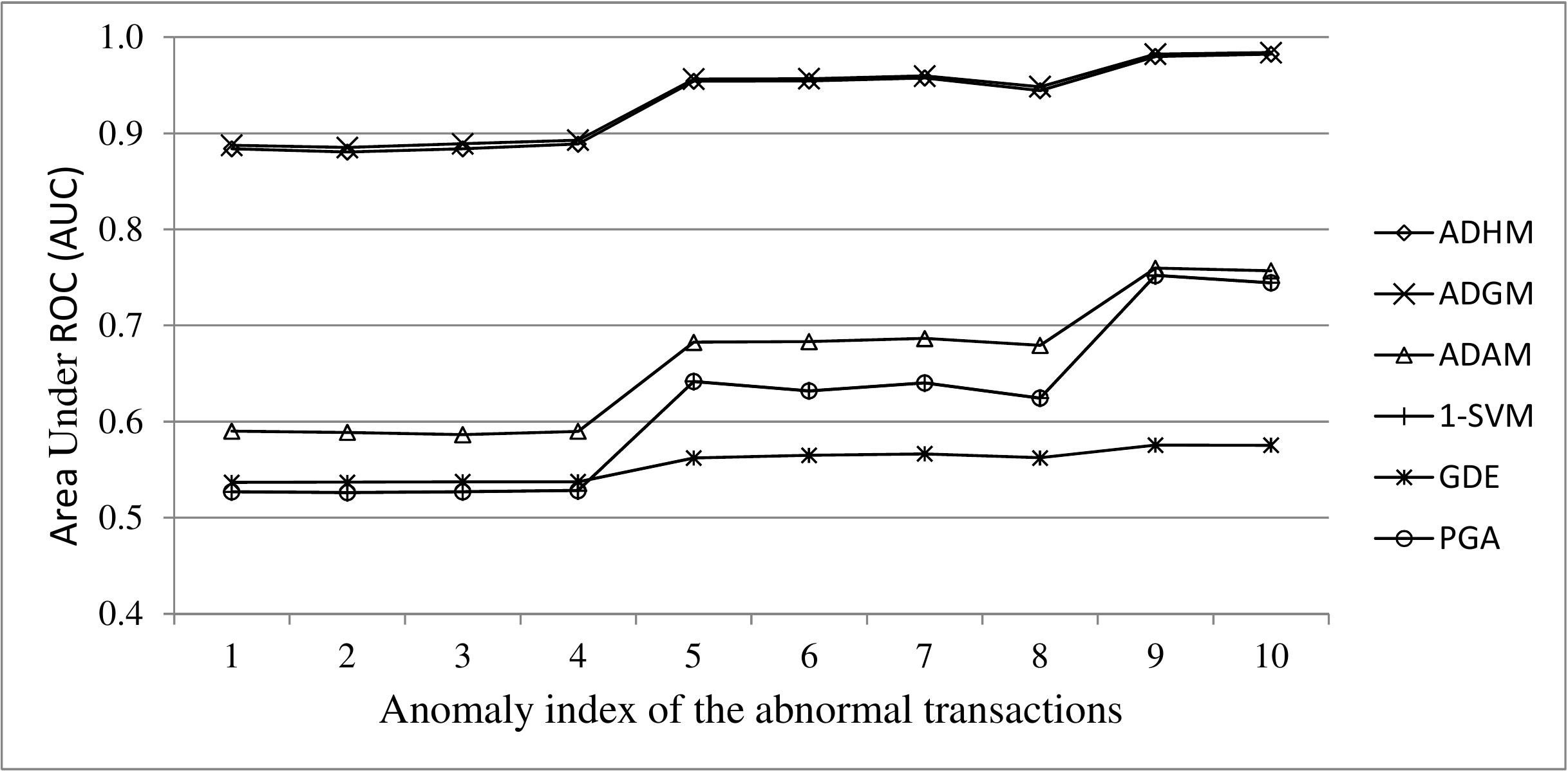}
		}
	\caption{AUC vs. XML anomaly index (\emph{Inventory dataset})}
	\label{fig:Inventory}
\end{figure}
%\vspace{-2mm}

The experimental results, presented in Figures \ref{fig:Insurance} and \ref{fig:Inventory}, show that the classification effectiveness of ADIFA-GM and ADIFA-HM is consistently better than the rest of the tested classifiers across all the tested anomaly indexes. Interestingly, if the classifiers where to be ranked according to their relative AUC performance, (i.e., the best classifier is ranked ``1'', the second best is ranked ``2'', and so on) the resulted ranking would be the same for every anomaly index. This result indicates that the dataset's anomaly index has no influence on the responsiveness of the classifiers to anomalies. Thus, the responsiveness must be an intrinsic feature of the inducers, which were used to construct the tested classifiers. We also observed that the multivariate classifiers (i.e., PGA, 1-SVM and GDE) did not show any improvement in the \emph{Insurance} dataset as the anomaly index increased. This, however, was not the case in the \emph{Inventory} dataset, which has a much lower dimensionality in comparison. This further supports our claim that anomaly classifiers, which are based on multivariate distance functions, are less suitable for high dimensional datasets, particularly, for XML anomaly detection.

\subsection{Learning Curves}
A learning curve reflects the relation between the efforts to learn and the classification performance of the trained classifier. Thus, a learning curve is an indication of the effectiveness of the learning algorithm. Additionally, since the learning curve indicates the rate in which the learning process converges, it can be used to determine the optimal amount of data for training (i.e., the optimal ``learning effort''). 

\begin{figure}[h]
	\centering
	\resizebox{1\linewidth}{!} {
		\includegraphics{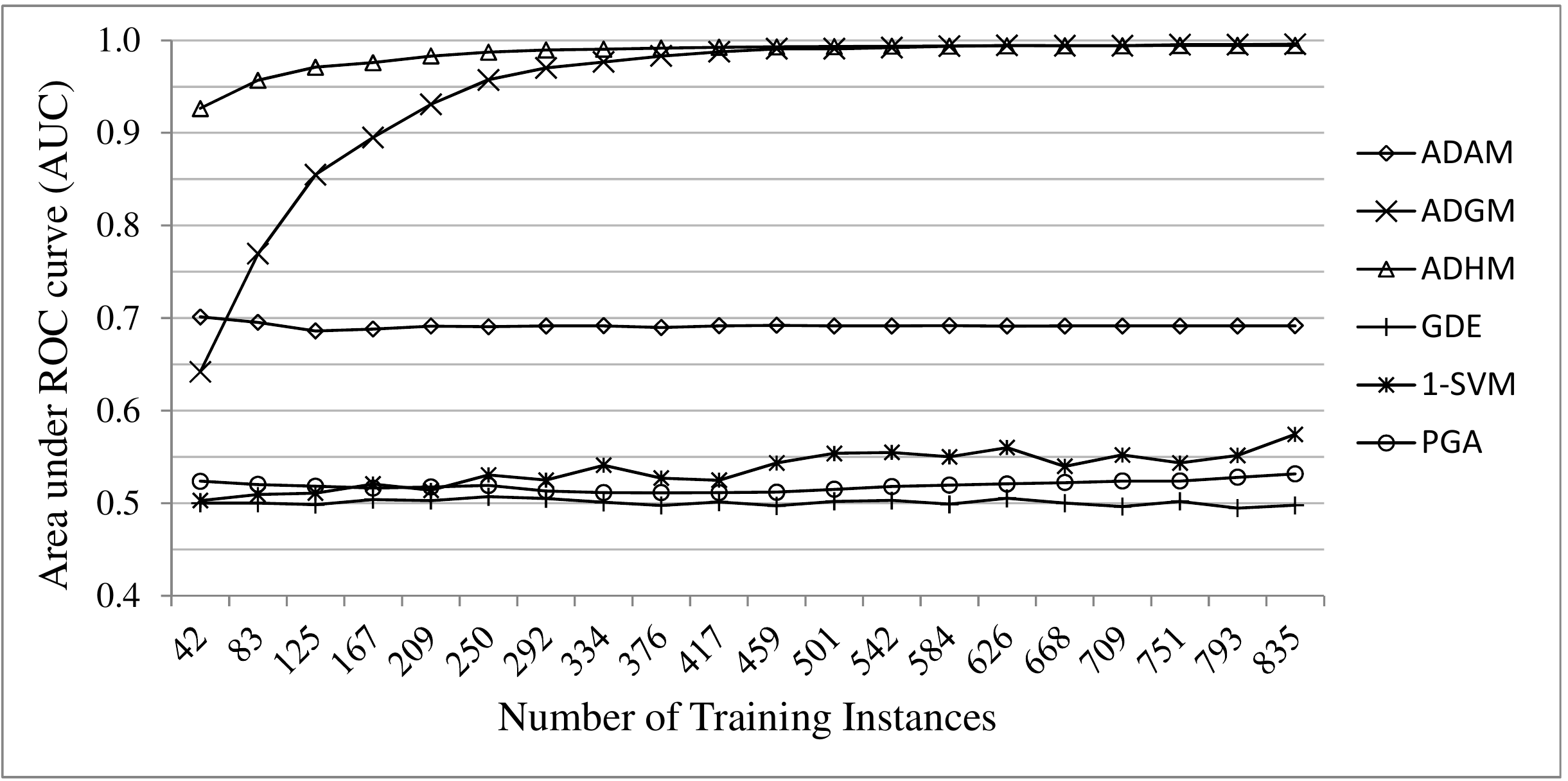}
		}
	\caption{Learning Curves for Insurance dataset}
	%\vspace{-3mm}
	\label{fig:LearningCurve_Insurance}
\end{figure}

\begin{figure}[h]
	\centering
		\resizebox{1\linewidth}{!} {
		\includegraphics{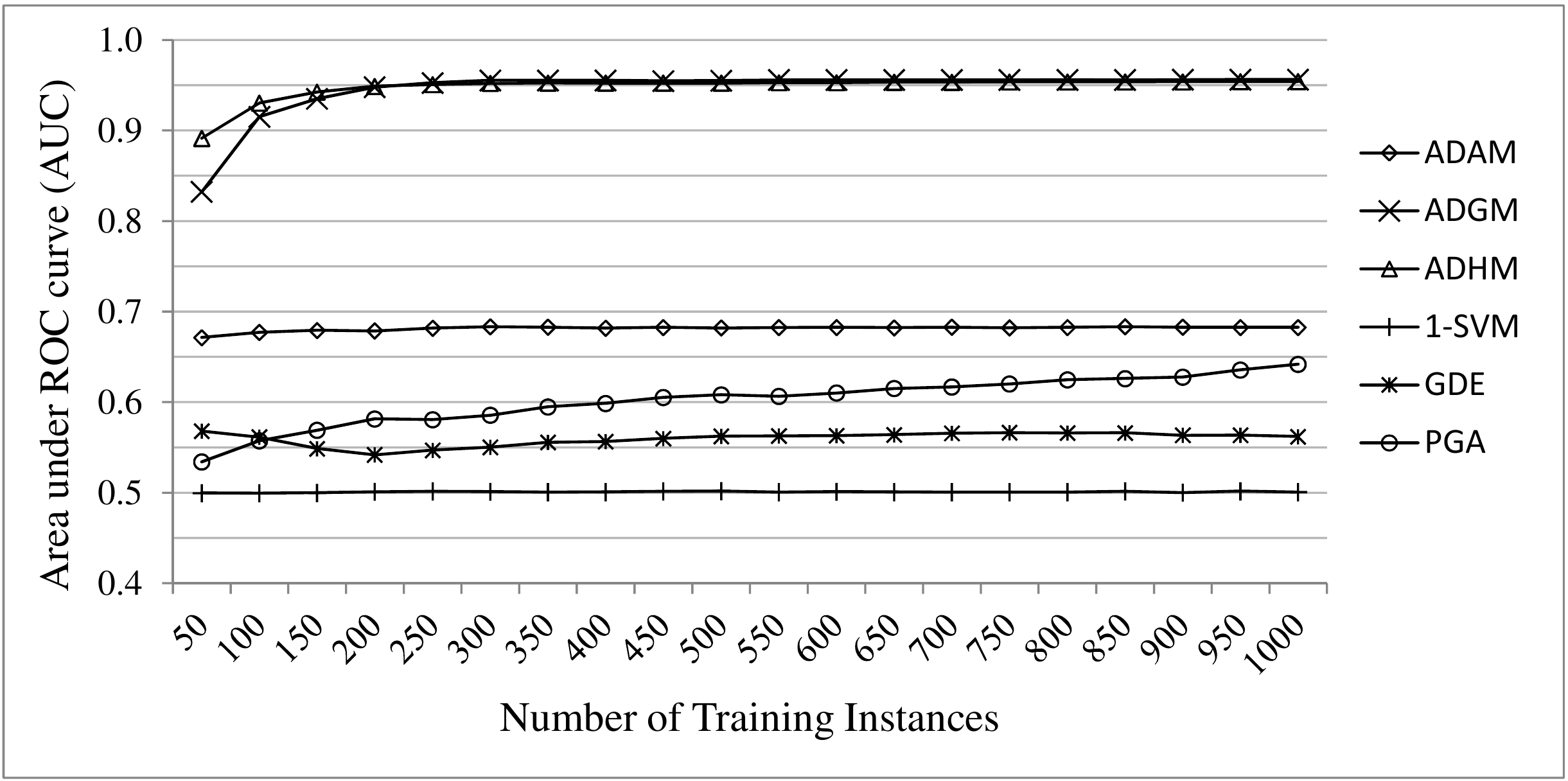}
		}
	\caption{Learning Curves for Inventory dataset}
	%\vspace{-3mm}
	\label{fig:LearningCurve_Inventory}
\end{figure}

\begin{figure}[h]
	\centering
		\resizebox{1\linewidth}{!} {
		\includegraphics{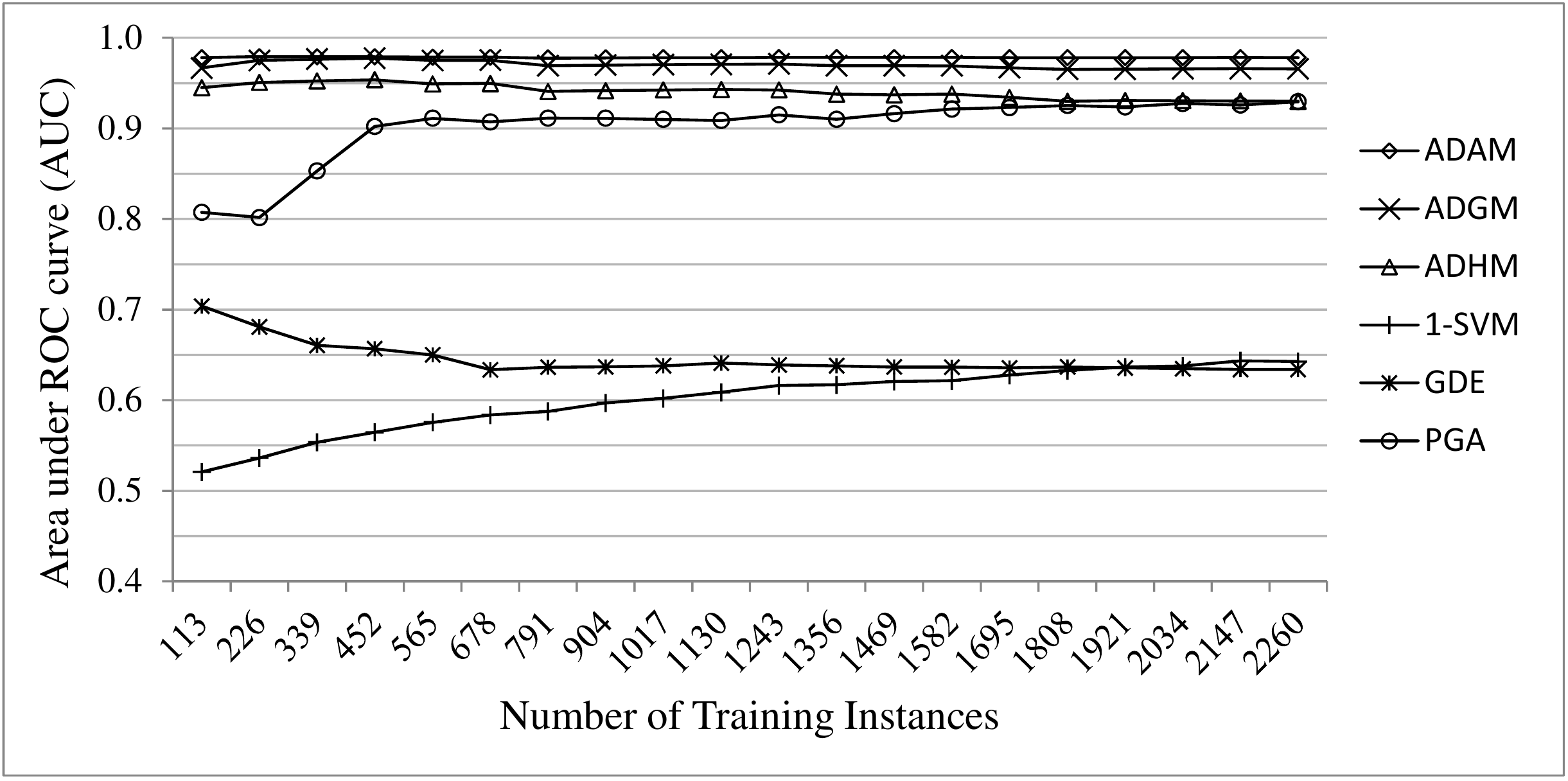}
		}
	\caption{Learning Curves for the ARP-D dataset}
	%\vspace{-3mm}
	\label{fig:LearningCurve_ARP}
\end{figure}

In the current experiment we derived ten new datasets from each of our original XML datasets. Each derived dataset contained a fix number of anomalous XML transactions, but a different amount of \emph{normal} instances, of which half were used for training the classifiers. In order to create the derived dataset, we applied the following process to each of the original XML transaction datasets. First, we divided the original dataset into two subsets: $D^-$ and $D^+$, where $D^-$ contained only the \emph{normal} XML transactions, while $D^+$ consisted of only the \emph{anomalous} XML transactions. Next, $D^-$ was randomly partitioned into ten equal sized groups: $D^{-}_1, \dots, D^{-}_{10}$. Then, the following datasets were constructed: $D_1$=$D^{+}\bigcup D^{-}_1$, $D_2$=$D_1 \bigcup D^{-}_2$, $\dots$, $D_{10}$=$D_9 \bigcup D^{-}_{10}$. Finally, we evaluated the tested learning algorithms on $D_1, \dots, D_{10}$. The results are depicted in Figures \ref{fig:LearningCurve_Insurance} - \ref{fig:LearningCurve_ARP}. 

%\begin{figure*}[tbh]
	%\begin{minipage}[l]{0.32\linewidth}
	%\centering
			%\includegraphics[scale=0.23]{LearningCurve_Insurance.pdf}
		%\caption{Learning Curves for \newline Insurance dataset}
		%\label{fig:LearningCurve_Insurance}
%\end{minipage}
%\hspace{0.2cm}
%\begin{minipage}[r]{0.32\linewidth}
	%\centering
	%\includegraphics[scale=0.23]{LearningCurve_Inventory.pdf}
	%\caption{Learning Curves for \newline Inventory dataset}
	%\label{fig:LearningCurve_Inventory}
%\end{minipage}
%\hspace{0.2cm}

%\begin{minipage}[r]{0.32\linewidth}
	%\centering
		%\includegraphics[scale=0.23]{LearningCurve_ARP-D1.pdf}
	%\caption{Learning Curves for the ARP-D dataset}
	%\label{fig:LearningCurve_ARP}
	%\end{minipage}
%\end{figure*}

The learning curves of the \emph{Insurance} and \emph{Inventory} datasets show that only ADIFA-GM and ADIFA-HM significantly improved their classification performance as the number of training instances increased. In contrast, the other classifiers improved by a much lower rate, if any. Among which, the 1-SVM classifier achieved the best performance improvement, although its learning curve was not monotonic.

The situation in the \emph{ARP-D} dataset was quite different from the two previously discussed datasets. 
%Most of the classifiers` performance did not improve when more training instances were available. The only exception here was the PGA, which kept improving slowly but steadily, as more training instances were available.
The classification performance of ADIFA-AM was on par with ADIFA-GM and better than that of ADIFA-HM. These three classifiers reached their optimal performance when they were applied to the smallest training-set, i.e., 113 instances. The GDE classifier has a negative learning curve, were its classification performance deteriorated as the number of training instances increased. This trend was somewhat restraint when the training set contained more than 400 instances. 

\subsection{Can ADIFA Generalize to Other Domains?}
In this section we examine whether the ADIFA algorithm can detect anomalies in other domains as well as it did within the XML transaction domain, i.e., does the multi-univariate approach of ADIFA work in other classification domains. 
To answer this question, in the following experiment we compare the performance of ADIFA with the performance of the anomaly detection algorithms in Table \ref{tab:ClassifiersSetup}, and also, with the Local Outlier Factor algorithm \cite{DBLP:conf/sigmod/BreunigKNS00} \footnote{The LOF algorithm was considered also for the XML-AD framework evaluation, but was ruled out due to the algorithm's inability to handle high dimensionality data, such as with the Insurance and Inventory dataset}.
The Local Outlier Factor (LOF) is a density-based anomaly detection algorithm, which is usually applied for unsupervised anomaly detection (i.e., when none of the training instances are labeled). In order to apply LOF in the one-class paradigm, we apply the following rule to the LOF output. 
We first use the LOF algorithm to calculate a local outlier factor value, $lof(x)$, for each instance $x$ in the training set, $X$. Let $LOF_{max}=\operatorname{max}_{x\in X} lof(x)$ be the highest calculated local outlier factor.
A test instance $x`$ is labeled as \emph{anomalous} if $LOF_{max}\leq lof(x`)$, where $lof(x`)$ is the calculated local outlier factor of the test instance $x`$. This rule is suitable for our purpose since $LOF_{max}$ is the maximal observed local outlier factor value in the entire training set, and therefore, approximately bounds the maximal acceptable local outlier factor value for \emph{normal} test instances.

We selected 30 popular datasets from the widely used UCI repository \cite{FrankAsuncion2010}. The datasets vary across characteristics such as the number of target classes, instances, input features, and feature types (nominal, numeric). In order to have only two classes in each dataset, we only selected the instances of the two most prominent classes. Similar to the previous experiment, only instances of a single class (the first of the defined classes), where used for training, while the instances of the second class were used strictly for evaluation. The results are displayed in Table \ref{tab:ResultTableForTheUCIDatasets} and their statistical significance is presented in Table \ref{tab:significance}.

\begin{table}[ht]
	\centering
	
	%\scriptsize
		\resizebox{1\linewidth}{!} {
		\begin{tabular}{@{}l@{\hspace{3.0pt}} c @{\hspace{3.0pt}} c@{\hspace{3.0pt}} c@{\hspace{3.0pt}}c@{\hspace{3.0pt}} c@{\hspace{3.0pt}} c@{\hspace{3.0pt}} c@{}}
		\hline
					& \multicolumn{6}{c}{Classifiers} \\ \cline{2-8}
		 Datasets & AD-AM & AD-GM & AD-HM & GDE & PGA & 1-SVM & LOF\\
		\hline
		Audiology 			& 0.865 (4) & 0.909 (2) & 0.938 (1) & 0.803 (5) & 0.522 (7) & 0.627 (6) & 0.883 (3) \\
		Balance-Scale 	& 0.946 (2) & 0.940 (3) & 0.921 (4) & 0.741 (5) & 0.652 (6) & 0.638 (7) & 0.968 (1) \\
		Breast-cancer 	& 0.498 (5) & 0.522 (3) & 0.537 (2) & 0.416 (6) & 0.508 (4) & 0.541 (1) & 0.394 (7) \\
		C.H. Disease 		& 0.803 (1) & 0.766 (2) & 0.707 (3) & 0.681 (4) & 0.601 (6) & 0.491 (7) & 0.671 (5) \\
		Credit-rating 	& 0.832 (1) & 0.832 (2) & 0.807 (3) & 0.668 (5) & 0.605 (6) & 0.500 (7) & 0.763 (4) \\
		Ecoli 					& 0.940 (4) & 0.963 (1) & 0.963 (2) & 0.819 (6) & 0.634 (7) & 0.879 (5) & 0.942 (3) \\
		Heart-statlog 	& 0.815 (1) & 0.781 (2) & 0.722 (3) & 0.701 (5) & 0.577 (6) & 0.492 (7) & 0.709 (4) \\
		Hepatitis 			& 0.811 (3) & 0.825 (2) & 0.831 (1) & 0.701 (5) & 0.606 (6) & 0.484 (7) & 0.714 (4) \\
		Horse-colic 		& 0.781 (2) & 0.794 (1) & 0.771 (3) & 0.537 (5) & 0.523 (6) & 0.512 (7) & 0.693 (4) \\
		H. Disease 			& 0.816 (1) & 0.786 (2) & 0.739 (4) & 0.761 (3) & 0.690 (5) & 0.503 (7) & 0.640 (6) \\
		Ionosphere 			& 0.799 (6) & 0.901 (3) & 0.955 (1) & 0.696 (7) & 0.895 (4) & 0.812 (5) & 0.926 (2) \\
		Iris 						& 0.998 (4) & 1.000 (1) & 1.000 (1) & 0.894 (7) & 0.980 (5) & 0.931 (6) & 1.000 (1) \\
		Labor 					& 0.943 (1) & 0.910 (2) & 0.873 (3) & 0.500 (7) & 0.611 (5) & 0.568 (6) & 0.856 (4) \\
		Lymphography 		& 0.810 (1) & 0.785 (3) & 0.756 (5) & 0.774 (4) & 0.606 (6) & 0.563 (7) & 0.803 (2) \\
		Mfeat 					& 0.985 (4) & 0.985 (3) & 0.995 (1) & 0.915 (5) & 0.765 (6) & 0.725 (7) & 0.990 (2) \\
		Mushroom 				& 0.925 (2) & 0.910 (3) & 0.888 (4) & 0.666 (5) & 0.628 (6) & 0.618 (7) & 0.995 (1) \\
		Page-blocks 		& 0.858 (4) & 0.909 (3) & 0.936 (2) & 0.684 (6) & 0.844 (5) & 0.583 (7) & 0.973 (1) \\
		Pen-Digits 			& 1.000 (2) & 1.000 (1) & 0.999 (3) & 0.826 (6) & 0.978 (5) & 0.749 (7) & 0.998 (4) \\
		Diabetes 				& 0.522 (2) & 0.512 (3) & 0.509 (4) & 0.447 (7) & 0.490 (5) & 0.583 (1) & 0.469 (6) \\
		Primary Tumor 	& 0.664 (1) & 0.587 (3) & 0.535 (7) & 0.564 (5) & 0.567 (4) & 0.547 (6) & 0.636 (2) \\
		Segment 				& 0.998 (1) & 0.995 (4) & 0.997 (2) & 0.849 (6) & 0.975 (5) & 0.824 (7) & 0.996 (3) \\
		Sick 						& 0.825 (1) & 0.823 (2) & 0.820 (3) & 0.705 (5) & 0.616 (6) & 0.517 (7) & 0.779 (4) \\
		Spam-base 			& 0.804 (3) & 0.837 (1) & 0.827 (2) & 0.607 (4) & 0.582 (5) & 0.555 (6) & 0.494 (7) \\
		Splice 					& 0.986 (3) & 0.997 (2) & 0.997 (1) & 0.575 (5) & 0.549 (6) & 0.484 (7) & 0.968 (4) \\
		Vehicle 				& 0.724 (4) & 0.757 (3) & 0.785 (2) & 0.679 (5) & 0.626 (6) & 0.482 (7) & 0.802 (1) \\
		Vote 						& 0.959 (3) & 0.972 (2) & 0.983 (1) & 0.721 (5) & 0.488 (7) & 0.757 (4) & 0.672 (6) \\
		Vowel 					& 0.695 (4) & 0.731 (3) & 0.758 (2) & 0.797 (1) & 0.627 (6) & 0.582 (7) & 0.680 (5) \\
		Waveform 				& 0.832 (4) & 0.849 (2) & 0.840 (3) & 0.627 (7) & 0.710 (6) & 0.831 (5) & 0.861 (1) \\
		W-Breast-Cancer & 0.991 (2) & 0.992 (1) & 0.984 (3) & 0.797 (6) & 0.923 (5) & 0.610 (7) & 0.969 (4) \\
		Mfeat-Factors 	& 0.990 (4) & 0.999 (1) & 0.999 (2) & 0.896 (6) & 0.974 (5) & 0.500 (7) & 0.994 (3) \\
		\hline
		\emph{Average} 	& 0.85 (2.67) & \myem{0.85 (2.2)} & 0.85 (2.6) & 0.70 (5.3) & 0.68 (5.6) & \mybf{0.62 (6.1)} & 0.81 (3.5) \\
		\hline		
		\end{tabular}}
		%\vspace{-2mm}
	\caption{Average AUC result for the UCI datasets. Inside the parenthesis is the AUC rank of the tested classifier. }
	\label{tab:ResultTableForTheUCIDatasets}
\end{table}

The results show that, on average, the three ADIFA variations perform better, than the multivariate classifiers: OC-GDE, OC-PGA, 1-SVM, and LOF. The non-parametric Bonferroni Dunn test shows that while there is no significant difference between the ADIFA variations, only the ADIA-GM is significantly better than all the multivariate classifiers. The comparable performance of the three ADIFA variations may be attributed to the low-dimensionality of the selected UCI datasets. Specifically, in low-dimensionality, the aggregation functions which are used by the different ADIFA variations, are highly correlated, leading to similar classifications.

\begin{table}[t]
	\centering
	\resizebox{1\linewidth}{!} {
		\begin{tabular}{@{}l c c c c c r}
		\hline
					& \multicolumn{6}{c}{Classifiers} \\ \cline{2-7}				
					& AD-AM & AD-GM & AD-HM & OC-GDE & OC-PGA & 1-SVM\\
			\hline
			AD-GM 	& $=$ &  		&  		&  		&  		& 		\\
			AD-HM 	& $=$ & $=$ &  		&  		&  		& 		\\
			OC-GDE 	& $+$ & $+$ & $+$ &  		&  		& 		\\
			OC-PGA 	& $+$ & $+$ & $+$ & $=$ &  		& 		\\
			1-SVM  	& $+$ & $+$ & $+$ & $=$ & $=$ & 		\\
			LOF   	& $=$ & $+$ & $=$ & $-$ & $-$ & $-$ \\
		\hline
	\end{tabular}}
	\caption{The significance of the difference between the classifiers using the AUC metric. Row $i$ and column $j$ contain: a `+' in case the method in the column is significantly better than the method in the row; a `-' indicates the opposite, and a `=' indicates that the difference is insignificant}
	\label{tab:significance}
\end{table}

\section{Conclusions And Future Work}
\label{conclusion}
This paper presented a new framework for detecting XML anomalies. Our experiments showed that the approach taken in XML-AD is very useful for detecting various types of anomalies, some of which originate from possible attacks on the structure and content of XML transaction documents. 

One of the foundational challenges we faced in this research was finding general and efficiently predictive XML features that could be extracted from any XML transaction. We developed an automatic feature extraction process in which both the XML content and structure features were addressed. 

A key feature of the proposed framework is its unique method for transforming complex XML features into fixed-length feature-vectors (i.e., instance flattening). This feature makes it possible to use general anomaly detection algorithms, which are readily available. The price for this XML features transformation was relatively low, both in computation time and in information-loss, since the most critical feature values (for deciding whether the instance is abnormal) were preserved during the flattening process.

Most of the prominent existing XML anomaly detection algorithms are based on the association-rules or multivariate distance function, which both perform poorly in high dimensions. We therefore proposed a new algorithm, ADIFA, that is comprised of multiple univariate models. Our evaluation demonstrated ADIFA's performance superiority over four related algorithms (1-SVM, OC-PGA, OC-GDE and LOF) both in detecting anomalies in XML transactions and in other domains.

Future work may include evolving our XML-AD framework towards a transaction filtering system, which, in addition to performing anomaly detection, could also prevent system attacks, i.e., a machine-learning based XML-firewall.

%\end{document}  % This is where a 'short' article might terminate

%ACKNOWLEDGMENTS are optional
%\section{Acknowledgments}
%We would like to thank god for creating such a marvel universe - keep up the hard work! b.t.w. we shell take it from here, see you later dude.
\bibliographystyle{abbrv}
\bibliography{XML-AD-Arxives}  % sigproc.bib is the name of the Bibliography in this case
% ACM needs 'a single self-contained file'!
%\balancecolumns

% That's all folks!
\end{document}